\documentclass{aa}  
\usepackage{graphicx,color}
\usepackage{txfonts}
\usepackage{amsmath}

\usepackage{natbib,twoopt}
\usepackage[hyphenbreaks]{breakurl}
\usepackage[breaklinks]{hyperref}  
\DeclareMathAlphabet\mathzapf       {T1}{pzc} {mb} {it}
\bibpunct{(}{)}{;}{a}{}{,}             
\definecolor{cobalt}{rgb}{0.06, 0.2, 0.65}
\hypersetup{
  colorlinks,
  citecolor=cobalt,
  linkcolor=cobalt,
  urlcolor=cobalt,
}
\makeatletter
  \newcommandtwoopt{\citeads}[3][][]{\href{http://adsabs.harvard.edu/abs/#3}%
    {\def\hyper@linkstart##1##2{}%
     \let\hyper@linkend\@empty\citealp[#1][#2]{#3}}}
  \newcommandtwoopt{\citepads}[3][][]{\href{http://adsabs.harvard.edu/abs/#3}%
    {\def\hyper@linkstart##1##2{}%
     \let\hyper@linkend\@empty\citep[#1][#2]{#3}}}
  \newcommandtwoopt{\citetads}[3][][]{\href{http://adsabs.harvard.edu/abs/#3}%
    {\def\hyper@linkstart##1##2{}%
     \let\hyper@linkend\@empty\citet[#1][#2]{#3}}}
  \newcommandtwoopt{\citeyearads}[3][][]%
    {\href{http://adsabs.harvard.edu/abs/#3}
    {\def\hyper@linkstart##1##2{}%
     \let\hyper@linkend\@empty\citeyear[#1][#2]{#3}}}
\makeatother

\def\xmm{{\it XMM-Newton}}
\def\meer{MeerKAT}

\newcommand{\cxo}{{\em Chandra}}
\newcommand{\swift}{{\em Swift}}
\newcommand{\fermi}{{\em Fermi}}
\newcommand{\nustar}{{\em NuSTAR}}

\newcommand{\srcfirst}{\object{CXOU\,J110926.4$-$650224}}
\newcommand{\src}{J1109}
\newcommand{\igrfirst}{\object{IGR\,J18245$-$2452}}
\newcommand{\igr}{J1824} 
\newcommand{\psrfirst}{\object{PSR\,J1023$+$0038}} 
\newcommand{\psr}{J1023}

\newcommand{\rxsfirst}{\object{1RXS\,J154439.4$-$112820}}
\newcommand{\rxs}{J1544}
\newcommand{\fglfirst}{\object{4FGL\,J0427.8$-$6704}}
\newcommand{\fgl}{J0427}
\newcommand{\parallax}{\ensuremath{\varpi}}
\newcommand{\be}{\begin{equation}}
\newcommand{\en}{\end{equation}}

\def\ltsima{$\; \buildrel < \over \sim \;$}
\def\lsim{\lower.5ex\hbox{\ltsima}}
\def\gtsima{$\; \buildrel $\geq$ \over \sim \;$}
\def\gsim{\lower.5ex\hbox{\gtsima}}

\def\arc{\mbox{$^{\prime\prime}$}}

\def\deg{\mbox{$^{\circ}$}}
\def\nh{\hbox{$N_{\rm H}$}}
\def\flux {\mbox{erg~cm$^{-2}$~s$^{-1}$}}
\def\lum {\mbox{erg~s$^{-1}$}}
\defcitealias{cotizelati19}{CZ19}

\begin{document} 

 \title{Simultaneous X-ray and radio observations of the transitional millisecond pulsar candidate \srcfirst}
 \subtitle{The discovery of a variable radio counterpart}
  \titlerunning{Simultaneous X-ray and radio observations of \srcfirst}
 \authorrunning{F. Coti Zelati et al.}

 \author{F.~Coti~Zelati\inst{1,2,3}
	     \and 
	     B.~Hugo\inst{4,5}
	     \and
	     D.~F.~Torres\inst{1,2,6}
	     \and
	     D.~de~Martino\inst{7}
	     \and
	     A.~Papitto\inst{8}
	     \and
	     D.~A.~H. Buckley\inst{9,10,11}
	     \and
	     T.~D.~Russell\inst{12}
	     \and	     
	     S.~Campana\inst{3}
	     \and	
	     R.~Van~Rooyen\inst{4} 
	     \and
	     E.~Bozzo\inst{13}
	     \and
	     C.~Ferrigno\inst{13}
	     \and
	     J.~Li\inst{14,15}
	     \and
	     S.~Migliari\inst{16,17}
	     \and
	     I.~Monageng\inst{11,9}
	     \and
	     N.~Rea\inst{1,2}     
	     \and
	     M.~Serylak\inst{4,18}
	     \and
	     B.~W.~Stappers\inst{19} 
	     \and
	     N.~Titus\inst{11,9}
 	     }

 \institute{Institute of Space Sciences (ICE, CSIC), Campus UAB, Carrer de Can Magrans s/n, E-08193, Barcelona, Spain\\
            \email{cotizelati@ice.csic.es}
       	   \and
           Institut d'Estudis Espacials de Catalunya (IEEC), Carrer Gran Capit\`a 2--4, E-08034 Barcelona, Spain
           \and
	   INAF--Osservatorio Astronomico di Brera, Via Bianchi 46, I-23807 Merate (LC), Italy
           \and
           South African Radio Astronomy Observatory, 2 Fir Street, Black River Park, Observatory, Cape Town 7925, South Africa
           \and
	   Department of Physics and Electronics, Rhodes University, PO Box 94, Grahamstown 6140, South Africa
	   \and
	   Instituci\'o Catalana de Recerca i Estudis Avan\c cats (ICREA), Passeig Llu\'is Companys 23, E-08010 Barcelona, Spain
	   \and
	   INAF -- Osservatorio Astronomico di Capodimonte, Salita Moiariello 16, I-80131 Napoli, Italy
	   \and
	   INAF -- Osservatorio Astronomico di Roma, via Frascati 33, I-00040 Monte Porzio Catone (Roma), Italy
	   \and
	   South African Astronomical Observatory, PO Box 9, 7935 Observatory, Cape Town, South Africa
	   \and
	   Department of Physics, University of the Free State, PO Box 339, Bloemfontein, 9300, South Africa
	   \and
	   Department of Astronomy, University of Cape Town, Private Bag X3, Rondebosch 7701, South Africa
	   \and
	   INAF--Istituto di Astrofisica Spaziale e Fisica Cosmica, Via U.\,La Malfa 153, I-90146 Palermo, Italy
	   \and
 	   Department of Astronomy, University of Geneva, Chemin d'Ecogia 16, 1290 Versoix, Switzerland
	   \and
	   CAS Key Laboratory for Research in Galaxies and Cosmology, Department of Astronomy, University of Science and Technology of China, No.96 JinZhai Road Baohe District, Hefei 230026, Anhui, China
	   \and
	   School of Astronomy and Space Science, University of Science and Technology of China, No.96 JinZhai Road Baohe District, Hefei 230026, Anhui, China
	   \and
 	   Aurora Technology BV for the European Space Agency, ESAC/ESA, Camino Bajo del Castillo s/n, Urb. Villafranca del Castillo, 28691 Villanueva de la Ca\~nada, Madrid, Spain
 	   \and
 	   Institute of Cosmos Sciences, University of Barcelona, Mart\'i i Franqu\`es 1, E-08028 Barcelona, Spain
	   \and
	   Department of Physics and Astronomy, University of the Western Cape, Bellville, Cape Town 7535, South Africa
	   \and
	   Jodrell Bank Centre for Astrophysics, Department of Physics and Astronomy, The University of Manchester, Alan Turing Building, Oxford Road, Manchester, M13 9PL, UK
 }

 \date{Received May 31, 2021; accepted September 23, 2021}

\abstract{
We present the results of simultaneous observations of the transitional millisecond pulsar (tMSP) candidate \srcfirst\ 
with the \xmm\ satellite and the \meer\ telescope. 
The source was found at an average X-ray luminosity of $L_{\rm X}$\,$\simeq$\,$7\times10^{33}$\,\lum\ 
over the 0.3--10\,keV band (assuming a distance of 4\,kpc) and displayed a peculiar variability pattern in the X-ray emission, switching between high, low and flaring modes on timescales of tens of seconds. A radio counterpart was detected at a significance of 7.9$\sigma$ with an average flux density of $\simeq$33\,$\mu$Jy at 1.28\,GHz. It showed variability over the course of hours and emitted a $\simeq$10-min long flare just a few minutes after a brief sequence of multiple X-ray flares. No clear evidence for a significant correlated or anticorrelated variability pattern was found between the X-ray and radio emissions over timescales of tens of minutes and longer.
\srcfirst\ was undetected at higher radio frequencies in subsequent observations performed with the Australia Telescope
Compact Array, when the source was 
still in the same X-ray sub-luminous state observed before, down to a flux density upper limit of 15\,$\mu$Jy at 7.25\,GHz (at 3$\sigma$).  
We compare the radio emission properties of \srcfirst\ with those observed in known and candidate tMSPs and
discuss physical scenarios that may account for its persistent and flaring radio emissions.
}
   

\maketitle

\section{Introduction}
\label{sec:intro}

Transitional millisecond pulsars (tMSPs) are weakly magnetized ($B$\,$\approx$\,10$^8$--10$^9$\,G) neutron stars (NSs) in binary 
systems that spin hundreds of times per second and swing between different emission regimes, depending on the mass transfer 
rate from their low-mass ($<$\,1\,$M_{\odot}$) non-degenerate companion star. When the mass transfer is low or even switches off, the tMSPs are detected as MSP binaries with low X-ray luminosities ($L_{\rm X}$\,$\lesssim$\,$10^{32}$~\lum) and with no evidence for the presence of an accretion disk (e.g. \citealt{archibald09,demartino20} and references therein). Their emission from the radio to the gamma-ray band is powered mostly by the pulsar rotational energy and an intrabinary shock that forms due to the interaction between the wind of relativistic particles ejected by the pulsar and matter transferred from the companion (for a review, see \citealt{harding21}). 
At higher mass transfer rates, the tMSPs are observed in a X-ray sub-luminous state ($L_{\rm X}$\,$\simeq$\,$10^{33}$--$10^{34}$~\lum) characterised 
by the presence of an accretion disk around the NS and by peculiar multiband properties (see below), probably connected with the presence of
an active rotation-powered MSP inhibiting accretion onto the NS (e.g. \citealt{ambrosino17,bogdanov18,papitto19,veledina19,campana19,jaodand21}). 
In one case, a month-long bright X-ray outburst ($L_{\rm X}$\,$\gtrsim$\,$10^{36}$~\lum) powered by mass accretion onto the NS surface was also 
observed \citep{papitto13}. The tMSPs can linger in the rotation-powered state or the X-ray sub-luminous state 
for decades, and perform a transition on timescales of weeks (or even shorter), following a change in the mass transfer 
rate from the companion \citep{stappers14,bassa14}. 
Three tMSPs are known to date (for reviews, see \citealt{campana18,papitto20}).

The tMSPs display enigmatic phenomenological properties in the X-ray sub-luminous state. These include: a unique variability pattern in the X-ray emission, which manifests in the form of repeated switches between three intensity levels (henceforth dubbed 
`high', `low' and `flaring' modes) on timescales of tens of seconds; gamma-ray emission detectable with the \fermi\ satellite up to GeV energies at a luminosity of $L_{\gamma}$\,$\simeq$\,$10^{33}$--$10^{34}$~\lum,
comparable to that in the X-ray band; flaring activity and flickering at optical and near infrared wavelengths reminiscent of the X-ray mode transitions; 
relatively bright, variable radio continuum emission with a flat to slightly inverted spectrum \citep{hill11,ferrigno14,deller15}. 
This latter emission has been interpreted in terms of partially self-absorbed synchrotron radiation from a population of relativistic electrons that are launched in the form of a weak steady, compact jet (e.g. \citealt{blandford79,blandford82}), possibly in the presence of a quickly rotating NS magnetosphere that propels away the in-flowing plasma \citep{papitto15,deller15,campana16,cotizelati18}. However, subsequent observations of the prototypical tMSP \psrfirst\ (henceforth \psr) revealed a clear anticorrelated variability pattern between the X-ray and radio emissions \citep{bogdanov18}. This indicates that compact jet radiation can hardly account for the whole radio emission and that additional short-lived ejection processes of different origin are likely to operate closer to the NS for $\approx$20\,\% of the time, causing the enhancements in the radio brightness and the switch to the low mode in the X-rays. These enhancements are probably caused by outflows of optically-thin plasma, but the physical mechanism that drives such events is currently not well understood (see \citealt{bogdanov18} for an extensive discussion; see also \citealt{baglio19}). It may be related to a rapid displacement (to a greater distance from the NS) of a termination shock that forms due to the interaction between the particle wind ejected by a constantly active rotation-powered  MSP and the in-flowing plasma \citep{papitto19}. Alternatively, it may be associated with a switch from a rotation-powered regime to a propeller regime at the transition from the high mode to the low mode \citep{veledina19}.

Recently, we have discovered that the source \srcfirst\ (henceforth \src) shows optical and high-energy emission properties 
closely resembling those of tMSPs in the X-ray sub-luminous state (\citealt{cotizelati19}; \citetalias{cotizelati19} 
hereafter): broad Balmer and Helium emission lines in the optical spectrum displaying variable profiles at different 
epochs, indicating the presence of an accretion disk in the system; a trimodal variability pattern in the X-ray emission around an average luminosity of 
$L_X\approx2\times10^{34}$\,\lum\ (assuming a distance of 4\,kpc); a spatial association with a
gamma-ray source listed in the preliminary \fermi/LAT 8-yr point source list
with a luminosity akin to those of the transitional MSPs in the accretion disk state. 
\src\ was undetected at radio wavelengths in shallow observations with the Australia Telescope 
Compact Array (ATCA) coordinated with a \nustar\ X-ray pointing, seemingly at odds with what has been seen in the 
tMSPs. However, the flux upper limits were consistent with the radio luminosity 
expected for tMSPs at similar X-ray luminosities \citep{gallo18}. 

This manuscript presents the results of new, simultaneous observations of \src\ with the \xmm\ 
satellite and the \meer\ telescope, complemented by subsequent simultaneous X-ray and radio observations 
with the \emph{Neil Gehrels Swift Observatory} and ATCA. We describe the observations and the data analysis in 
Section~\ref{data} and report the results in Section~\ref{results}. A discussion and conclusions follow in Sections~\ref{discussion} and ~\ref{conclusions}.

\section{Observations and data processing}
\label{data}

Table~\ref{tab:log} reports the journal of the observations. In the following, we give details on the 
two campaigns, and describe the data processing and analysis. We adopt the following position for \src: R.A. = 11$^\mathrm{h}$09$^\mathrm{m}$26$\fs$40, decl. = --65$^{\circ}$02$^{\prime}$24$\farcs$80 (J2000.0\footnote{This was derived by applying a linear correction to the position  quoted in the early version of the third data release (EDR3) from the {\it Gaia} space telescope (which is given at the reference epoch J2016.0), based on the source proper motion, $\mu_{{\rm RA}}=1.17\pm0.96$\,mas\,yr$^{-1}$, $\mu_{{\rm decl}}=-1.60\pm0.67$\,mas\,yr$^{-1}$, and accounting for uncertainties in the absolute {\it Gaia} positions \citep{gaia2021}. The 1$\sigma$ uncertainties on the position so evaluated are $\sigma_{{\rm RA}}\simeq14$\,mas, $\sigma_{{\rm decl}}\simeq11$\,mas.}).

\begin{table}
\scriptsize
\caption{
\label{tab:log}
Observation log in 2019.}
\centering
\begin{tabular}{lcc}
\hline\hline
Telescope					&Start -- End time	&Exposure \\
						&Mmm DD hh:mm:ss (UTC)			& (ks) \\
\hline
\xmm/EPIC MOS1 			& Jun 14 18:46:04 -- Jun 15 02:26:26 & 26.8  \\
\xmm/EPIC MOS2 			& Jun 14 18:46:25 -- Jun 15 02:26:31 & 26.8  \\
\xmm/EPIC pn 				& Jun 14 19:23:42 -- Jun 15 02:30:37 & 24.9  \\
\xmm/OM\tablefootmark{a}	& Jun 14 18:54:26 -- Jun 15 02:29:47 & 25.5  \\
\meer					    & Jun 14 15:31:32 -- Jun 14 23:23:45 & 28.4  \\
\hline
\xmm/EPIC MOS1 			& Jun 15 14:24:17 -- Jun 16 00:51:19 & 36.5  \\
\xmm/EPIC MOS2 			& Jun 15 14:24:40 -- Jun 16 00:51:25 & 36.5  \\
\xmm/EPIC pn 				& Jun 15 15:01:55 -- Jun 16 00:51:39 & 34.4  \\
\xmm/OM\tablefootmark{b}   & Jun 15 16:56:13 -- Jun 16 00:24:40 & 25.1 	 \\
\meer					    & Jun 15 14:44:26 -- Jun 15 22:31:34 & 28.1  \\
\hline
\hline
\swift/XRT					& Sep 6 21:01:34 -- Sep 6 21:29:56 & 1.7  \\
\swift/XRT					& Sep 6 22:37:03 -- Sep 6 23:04:56 & 1.7  \\
\swift/XRT					& Sep 7 00:14:23 -- Sep 7 00:17:53 & 0.2  \\
ATCA					    & Sep 6 22:36:30 -- Sep 7 07:09:20 & 30.8 \\
\hline
\end{tabular}
\newline
{\bf Notes.}
\tablefoottext{a}{Seven exposures were acquired, with lengths ranging from 1.2 to 4.3\,ks.}
\tablefoottext{b}{The OM observation started about 8\,ks after the EPIC observation, due to an instrument anomaly. Seven exposures were acquired, with 
lengths from 1.2 to 4.0\,ks.}
\end{table}

\subsection{Observations in June 2019}

\subsubsection{\xmm}
\label{xmm}

The European Photon Imaging Cameras (EPIC; \citealt{struder01,turner01}) and the Optical/UV Monitor Telescope (OM; 
\citealt{mason01}) on board \xmm\ observed \src\ twice between 2019 June 14 and 16. The EPIC-MOS cameras were 
configured in small window mode, the EPIC-pn detector was set in fast timing mode and the OM operated in the fast window mode in white light. The data were processed and analyzed using the Science Analysis Software (\texttt{SAS} v.~19).

We converted the X-ray photons arrival times from the Terrestrial Time (TT) standard to the Coordinated Universal Time (UTC) standard. 
We discarded data acquired over the first 10\,ks of the first observation, owing to strong 
contamination from background flares, and retained subsequent events within the 0.3--10\,keV energy range for the following analysis. 
For the MOS cameras, we extracted the source photons from a circle of radius 30\arc\ centred on the source 
position and the background photons from a circle of radius 60\arc\ 
located on one of the outer CCDs. For the pn, we collected the source photons from a 10-pixel-wide strip oriented along the readout direction of the CCD ($32 < {\rm RAW}$-${\rm X} \leq 42$)
and the background photons from a 3-pixel-wide region far from the source ($3 < {\rm RAW}$-${\rm X} \leq 5$). \src\ was detected at the following net count rates: 0.186$\pm$0.003 counts\,s$^{-1}$ (June 14) and 0.159$\pm$0.002 counts\,s$^{-1}$ (June 15) with the MOS1; 0.199$\pm$0.003 counts\,s$^{-1}$ (June 14) and 0.161$\pm$0.002 counts\,s$^{-1}$ (June 15) with the MOS2; 0.580$\pm$0.009 counts\,s$^{-1}$ (June 14) and 0.498$\pm$0.005 counts\,s$^{-1}$ (June 15) with the pn. We 
extracted the source light curve by combining the background-subtracted time series acquired over the time intervals 
covered by the three EPIC cameras simultaneously in the 0.3--10\,keV energy range. We extracted the spectra retaining single to quadruple pixel events for the 
MOS, and single and double pixel events with energy $\geq$\,0.7\,keV for the pn\footnote{\label{fn:note1}\url{https://xmmweb.esac.esa.int/docs/documents/CAL-TN-0018.pdf}}. 

\src\ was detected by the OM only during the last six images in the second pointing at net count rates in the range $\approx$0.4--0.8\,counts\,s$^{-1}$.
A reliable estimate of the local background level in the OM images where the source was undetected is complicated owing to the presence of a time-variable flux excess at the source position caused by a nearby contaminating extended scattered light feature\footnote{\xmm\ helpdesk private communication.}.
Based on the mean background level over the whole field of view in the single images, we can estimate the following conservative upper limits on the count rates of \src: $\simeq$0.15\,counts\,s$^{-1}$ for each 4-ks image and $\simeq$0.3\,counts\,s$^{-1}$ for the 1.2-ks image acquired towards the end of the first observing run (at 3$\sigma$). We stress, however, that these values are very likely to underestimate the true background level at the source position.

\subsubsection{\emph{\meer}}
\label{meer}
\meer\ \citep{jonas16,camilo18,mauch20} observed \src\ twice on 2019 June 14 and 15 using 61 and 60 antennas, respectively. 
The total exposures were $\sim$7.9 and $\sim$7.8\,hr for the first and the second track, respectively (including overheads). The data were 
recorded at a central frequency of 1.284\,GHz with a total bandwidth of 856\,MHz split into 4096 frequency channels and an integration time of 8\,s. 
The source PKS\,1934$-$638 \citep{reynolds94} was used for bandpass and flux calibration. The source 
J0906$-$6829 (R.A. = 09$^\mathrm{h}$06$^\mathrm{m}$52$\fs$23, decl. = --68$^{\circ}$29$^{\prime}$39$\farcs$90, J2000) was used for gain calibration during the second epoch.

We flagged the data by applying a dilated hard static mask to remove interference from mobile telecommunications and navigational satellite transmission that contaminate the passband of the \meer\ $L$-band wideband system, as well as by applying further manual flagging. This resulted in a flagging in excess of 70\% for both observations. The flux scale and temporal variability calibrations for the second observation were then bootstrapped to the first observation using a model of the target field derived from self-calibration of the second epoch, so as to minimize relative errors in the flux scales between the two observations. 

We then corrected for the residual offset in the average flux scales of a cross-matched selection of compact AGNs detected at a significance $\gtrsim$50$\sigma$ on the time-averaged continuum maps of the two epochs ($1\sigma\approx 5$\,$\mu$Jy for both epochs). Sources are classified as compact by measuring their peak-to-integrated flux ratio using the Python Blob Detector and Source Finder (\textsc{pyBDSF}; \citealt{mohan15}). A ratio close to unity for a source detected at high significance indicates that the source is unresolved, thus largely not subject to slight changes in the Point Spread Function between epochs. This ensured a relative error in the flux scales below the 1\% level between the two epochs. The error on the absolute flux scale was quantified by transfer calibration onto our backup primary calibrator PKS\,B0407$-$65, and was found to be at most 7\%.

The data were self-calibrated using the \textsc{DDFacet} imager \citep{tasse18} and \textsc{Cubical} calibration \citep{kenyon18} software packages. We used the `Briggs' weighting scheme \citep{briggs95} with a robustness parameter of -0.3\footnote{This value was chosen so as to achieve a nearly optimal sensitivity in angular resolution and at the same time to reduce the sidelobes of the MeerKAT point-spread function.} to create multi-frequency synthesis maps of untapered \meer\ uv-coverages. This resulted in maps with synthesised beams of $7.38\arc \times 6.10\arc$ for the first epoch and $6.70\arc \times 5.91\arc$ for the second epoch. The field is dominated by off-axis direction-dependent calibration artefacts (in the form of radial stripes and rings), which prompted us to perform direction-dependent calibration and peeling after transfer calibration (see Appendix~\ref{sec:rime}).

After self-calibration and peeling successfully removed the sidelobe interference contaminating the target position, we subtracted the flux of the remaining AGN population in a field of view of size $2.5\deg\times2.5\deg$ using a 10$\sigma$ mask, explicitly excluding the target position. This allowed us to create snapshot images via the \texttt{clean} imaging task in the Common Astronomy Software Application (\textsc{casa}; \citealt{mcmullin07}) to synthesise dirty maps of the residuals centred on the target position, with integration times of 1\,min. These maps could be stacked further to obtain exposures of variable duration.

We also made shallowly deconvolved images of the data after self-calibration and peeling to ensure that the light curves extracted from these residual images were not subject to errors stemming from temporal errors in the calibrated flux scales. We picked several compact AGNs detected at high significance close to the target position and extracted average flux light curves to catch instances of substantial calibration errors in the direction-independent electronic gains. We found that the temporal relative errors did not exceed 6\% during the two observations. 

\begin{figure}
\begin{center}
\includegraphics[width=.45\textwidth]{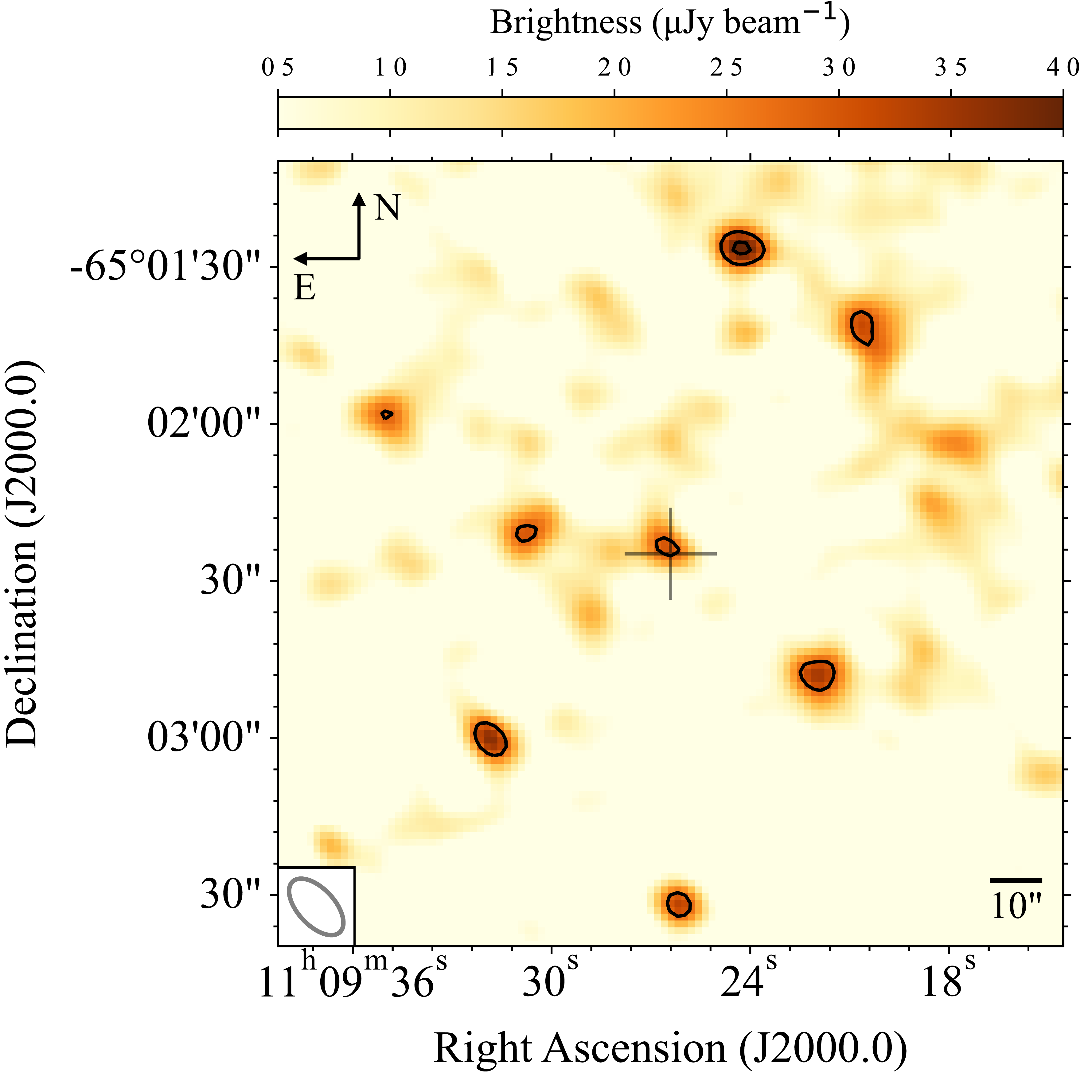}
\caption{1.284-GHz \meer\ image of the sky region around \src. The image was obtained from a stack of the data acquired on 2019 June 14 and 15 and its size is $2.5^{\prime}\times2.5^{\prime}$ (north is up and east to the left). The sky region is color-coded according to the values of the brightness at each position, as indicated by the color bar on the top. The rms noise level in this image is 4.2\,$\mu$Jy/beam. Contour levels are drawn at intervals of $(2^{n/2})\times30$\,$\mu$Jy/beam, where $n$ = 0, 1. The position of the optical counterpart of \src\ (see Section~\ref{xmm}) is marked using a grey cross at the centre of the image. The grey ellipse in the frame in the bottom-left corner represents the shape and size of the synthesised beam, where the major axis is 13.2\arc, the minor axis is 7.0\arc, and the major axis is located at a position angle of 43.0$^{\circ}$ measured from north towards east.} 
\label{fig:meer_image}
\end{center}
\end{figure}

Fig.\,\ref{fig:meer_image} shows the final image of the field around \src\ extracted from the stack of all the data. We measured a peak flux density of 33$\pm$4\,$\mu$Jy at 1.284\,GHz at the position of \src\ in this image, corresponding to a detection significance of $\simeq$7.9$\sigma$. The estimated peak-to-integrated flux ratio, $\simeq$1.8, is indicative of compact emission. We fit for the position of this emission using \textsc{pyBDSF} adopting an island cutoff of 3$\sigma$ and a confidence cutoff of 5$\sigma$, and derived the following position: 
R.A. = 11$^\mathrm{h}$09$^\mathrm{m}$26$\fs$53 $\pm$ 1$\fs$15, decl. = --65$^{\circ}$02$^{\prime}$23$\farcs$65 $\pm$ 0$\farcs$93 (J2000.0). This is compatible with the position derived from {\it Gaia} (and also from \cxo\ X-ray data; see \citetalias{cotizelati19}) within the uncertainties.

\subsection{Observations in September 2019}

\subsubsection{Swift \emph{XRT}}

The \swift\ X-ray Telescope (XRT; \citealt{burrows05}) observed \src\ in the photon counting mode starting 
on 2019 September 6 at 21:01:34 UTC. The observation was split into three snapshots, resulting in a total elapsed time of $\simeq$11.8\,ks. The second and the third snapshots were entirely covered by the ATCA 
observations (see Table~\ref{tab:log}). The data were processed with standard screening criteria. \src\ was detected at a net count rate of 
0.040$\pm$0.003 counts\,s$^{-1}$ over the energy range 0.3--10\,keV, corresponding to a signal-to-noise ratio of S/N$\simeq$10.5. The net count rates in the single snapshots were 0.043$\pm$0.005, 0.039$\pm$0.005 and 0.03$\pm$0.01 counts\,s$^{-1}$ (0.3--10\,keV), that is, consistent with each other within the uncertainties. Using the whole data set, we extracted the spectrum of \src\ from a circle of radius 47.2\arc\ centred on the target and the background spectrum from a source-free annulus with radii of 94.4\arc\  and 188.8\arc, also centred on the target.

\subsubsection{\emph{ATCA}}
\label{sec:atca}

We observed the field of \src\ with the ATCA on 2019 September 6--7, when the telescope was in an extended 6\,km (6C\footnote{\url{https://www.narrabri.atnf.csiro.au/operations/array_configurations/configurations.html}}) configuration. The observations were recorded simultaneously at central frequencies of 5.5 and 9\,GHz, with 2\,GHz of bandwidth at each central frequency. PKS\,1934$-$638 \citep{reynolds94} was used for primary bandpass and flux calibration, while PKS\,1059$-$63 was used for phase calibration. The data were edited for radio frequency interference (RFI) and instrumental issues, calibrated, and imaged following standard procedures in the CASA package. Imaging was carried out with a range of weightings (Briggs robustness parameters of 0, 1, and 2) in an attempt to maximise the image sensitivity and minimise the minor noise effects from a nearby bright source. 

\src\ was not detected in these observations, with 3-$\sigma$ upper-limits of 18\,$\mu$Jy at both 5.5 and 9\,GHz. To maximise the image sensitivity, we stacked both bands together, which also resulted in a non-detection with a 3-$\sigma$ upper limit of 15\,$\mu$Jy (centred at 7.25\,GHz). Stacking the ATCA images acquired in 2019 with those taken in 2018 (see \citetalias{cotizelati19}) resulted in our most sensitive image, but also provided a non-detection, yielding a 3-$\sigma$ upper-limit of 12\,$\mu$Jy at 7.25\,GHz.

\begin{figure*}
\centering
\includegraphics[width=0.95\textwidth]{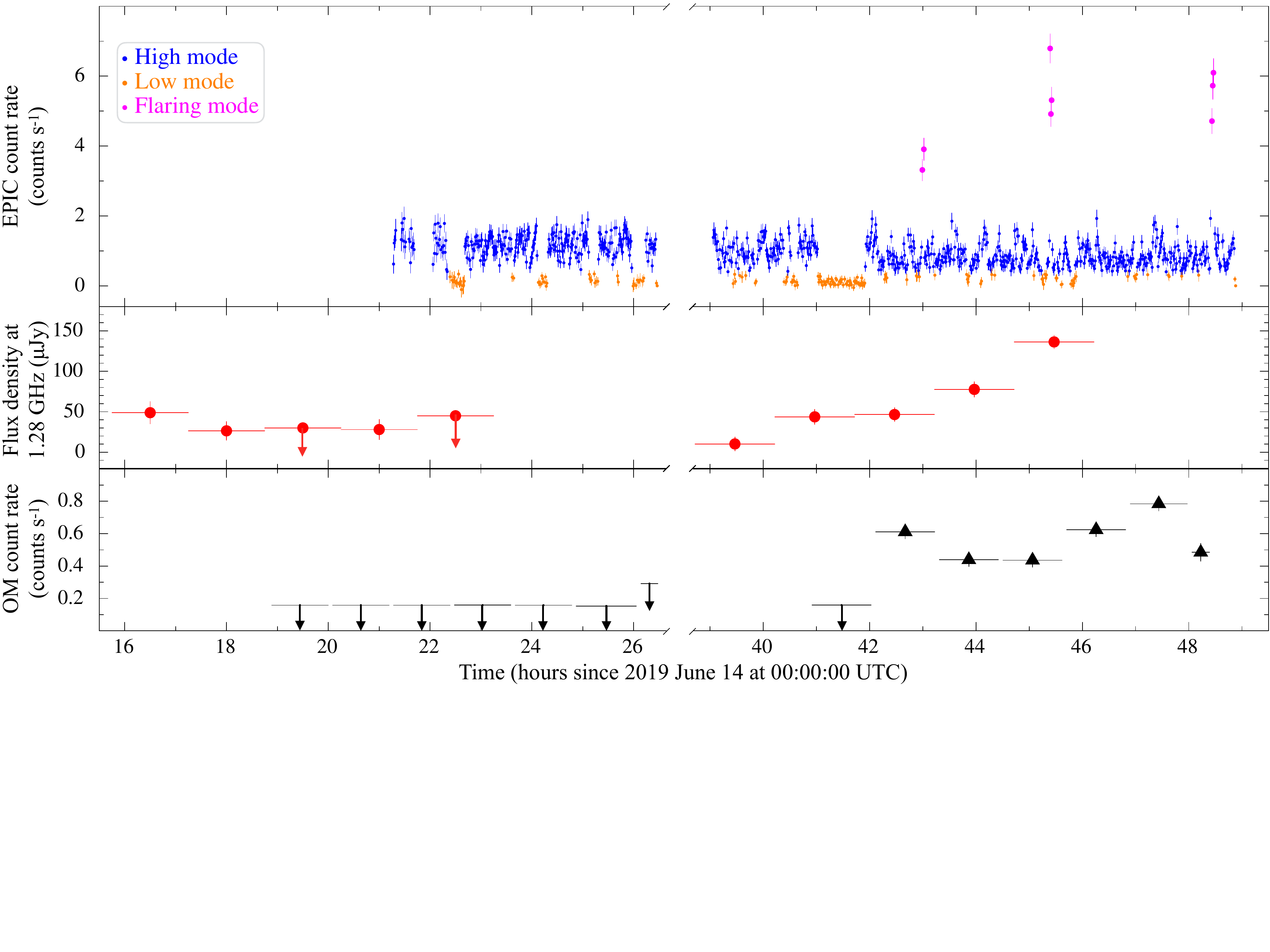}
\vspace{-3.5cm}
\caption{Multi-band light curves of \src\ on 2019 June 14--16. The X-ray light curve (top) is in the 0.3--10\,keV energy band and is binned at 50\,s. The flaring, high 
and low modes intervals are depicted in magenta, blue and orange, respectively (see the text for their definition). The radio light
curve (middle) is in the $L$ band (central frequency of 1.284\,GHz) and is binned at 1.5\,hr. The optical light curve (bottom) is in white light and was extracted using the net count rates measured in the single images (of lengths of 4 or 1.2\,ks). Upper limits are reported for the images where the source was undetected. In all panels, error bars represent 1$\sigma$ uncertainties, while upper limits are 
given at a confidence level of 3$\sigma$.}
\label{fig:lc}
\end{figure*}

\begin{figure}
\centering
\includegraphics[width=0.52\textwidth]{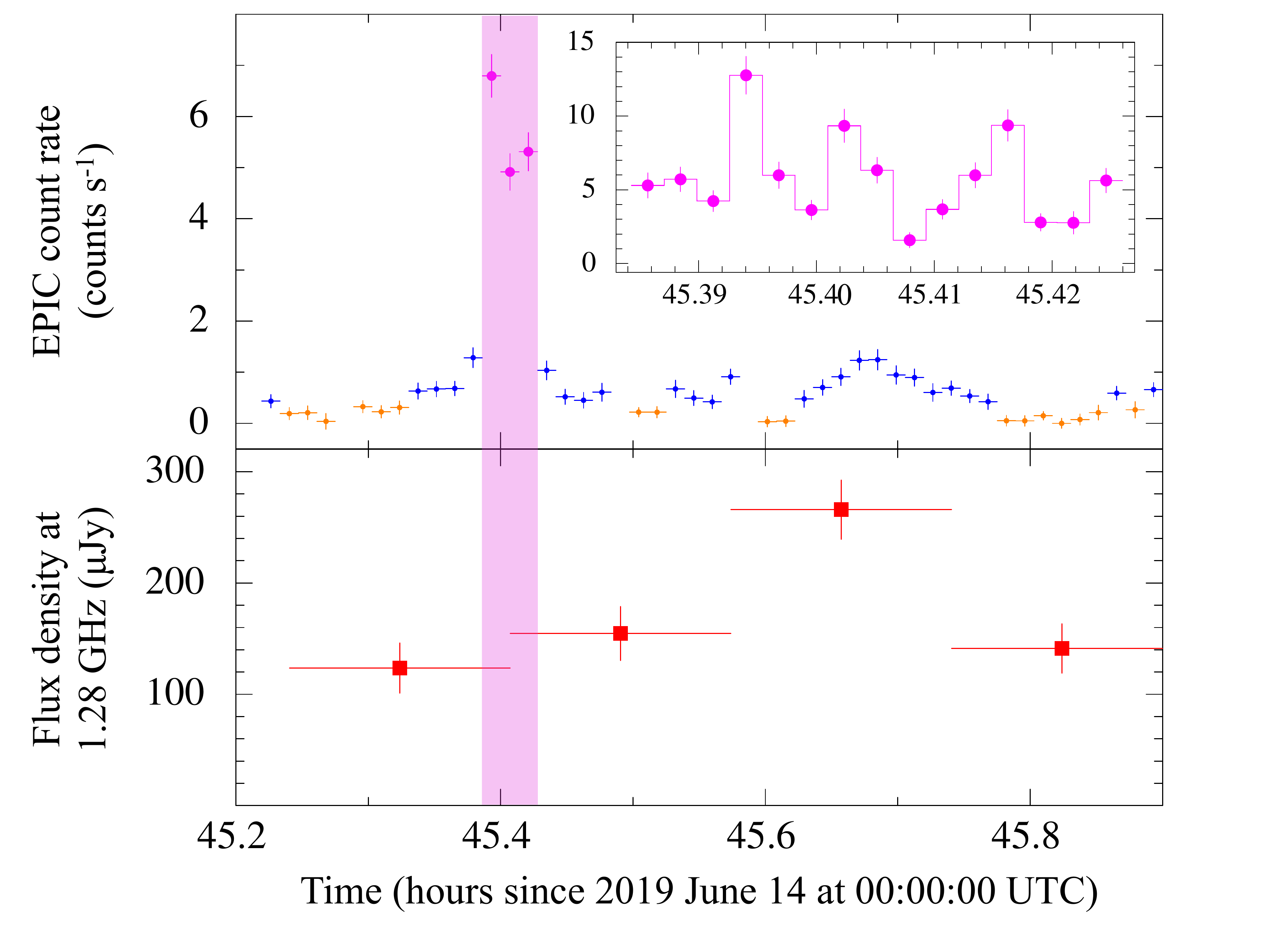}
\vspace{-0.6cm}
\caption{Zoom of the light curves around the epoch of the 
brightest X-ray flare (top; time bin of 50\,s) and the radio flare (bottom; time bin of 10\,min). The X-ray modes are color-coded as in Fig.\,\ref{fig:lc}. The shaded area in magenta marks the epoch of the X-ray flare. The inset in the top panel shows a further zoom of the X-ray light curve at the epoch of the X-ray flare and is binned at 10\,s. In all panels, error bars represent 1$\sigma$ uncertainties.}
\label{fig:lc_zoom}
\end{figure}

\section{Results}
\label{results}

\subsection{Multi-band variability}

The top panel of Fig.\,\ref{fig:lc} shows the X-ray light curve of \src. It displays a trimodal variability pattern that is remarkably akin to that observed in previous 
observations \citepalias{cotizelati19}, with the switches between the different modes occurring on timescales as short as tens of seconds. Here, we single out the different 
mode time intervals adopting the following thresholds for the net X-ray count rates (CR) measured in time bins of 50\,s: $0.4$~counts~s$^{-1}$ $\leq{\rm CR} 
< 2.0$~counts~s$^{-1}$ for the high mode; ${\rm CR}<0.4$~counts~s$^{-1}$ for the low mode; ${\rm CR}\geq2.4$~counts~s$^{-1}$ 
for the flaring mode. 
At some point during the second observation, the source was caught in the low mode for a duration of about 50\,min -- by far the longest period of low mode hitherto observed for this source. Later in the observation, the source underwent 
three short-lived flaring episodes. The count rate registered at the peak of the second event, $\simeq$\,7\,counts\,s$^{-1}$, 
was larger than those measured for the brightest flares in the previous \xmm\ observation in 2018 by a factor of $>$\,2, making this 
episode the X-ray flare with the highest net count rate at peak so far detected from \src. The corresponding light curve displayed a rather complex morphology, with at least three structured peaks detected within $\simeq$150\,s on timescales as short as 10\,s (see the inset in Fig.\,\ref{fig:lc_zoom}). 

We fit an absorbed power law (PL) model to the background-subtracted EPIC spectra of \src\ from the two observations jointly 
using the \texttt{Xspec} software (v.~12.12.0; \citealt{arnaud96}) to estimate the source average X-ray flux. The photoelectric absorption 
by the interstellar medium was described using the Tuebingen-Boulder model \citep{wilms00}, and the absorption column density was 
held fixed to $\nh=5.3\times10^{21}$~cm$^{-2}$ in the fits (see \citetalias{cotizelati19}). A renormalisation factor was included in the fits 
to account for intercalibration uncertainties across the EPIC cameras, yielding mismatches of $<$10\,\%. Results are reported in Table\,\ref{tab:flux}. The averaged unabsorbed flux, 
$F_{\rm X}$\,$\simeq$\,$3.5\times10^{-12}$ \flux\ (0.3--10\,keV) translates into a luminosity of $L_{\rm X}$\,$\simeq$\,$7\times10^{33}\,d_4^2$\,\lum\ 
(0.3--10\,keV), where $d_4$ is the source distance in units of 4\,kpc and isotropic emission is assumed (see Section~\ref{lxlradio} for a discussion on the source distance). We also evaluated the X-ray flux 
in each of the modes via a count rate-resolved spectral analysis of the data (see again Table\,\ref{tab:flux}).

\begin{table}
\footnotesize
\caption{
\label{tab:flux}
X-ray spectral properties of \src\ on 2019 June 14--16.}
\centering
\begin{tabular}{cccc}
\hline\hline
Emission state\tablefootmark{a}		&$\Gamma$\tablefootmark{b}	&$F_{\rm X}$\tablefootmark{c} &$\chi^2_{{\rm red}}$ (dof)  \\
				&						&(10$^{-12}$~\flux)	 \\
\hline
Average			& $1.60\pm0.01$			& $3.53\pm0.03$   				& 1.07 (209)	 	\\
\hline
High mode 		& $1.60\pm0.01$			& $3.91\pm0.04$  				& 1.12 (234) \\
Low mode  		& $1.5\pm0.1$				& $0.60\pm0.03$    				& 0.59 (34) \\
Flaring mode  		& $1.76\pm0.05$			& $19.5\pm0.5$    				& 0.95 (50) \\
\hline
\end{tabular}
{\bf Notes.}
\tablefoottext{a}{The spectra were fitted using an absorbed PL model with the absorption column density held fixed to $\nh=5.3\times10^{21}$~cm$^{-2}$.}
\tablefoottext{b}{Best-fitting photon index for the absorbed PL model.}
\tablefoottext{c}{Unabsorbed flux over the 0.3--10\,keV energy band.}
\end{table}

The faintness of the radio counterpart prevented us to probe the presence of variability in the radio emission detected by \meer\ on the same timescale of the 
X-ray mode switching. Nonetheless, we investigated possible changes by computing
the flux density in images extracted over time intervals of
variable length within the range 10\,min -- 2.5\,hr. The time intervals
of 1.5\,hr provide the best trade-off between time resolution
and S/N in order to characterize variability over the time span covered by our observations. This light
curve, shown in the middle panel of Fig.\,\ref{fig:lc}, reveals gradual variations
in the flux by a factor of $\gtrsim$5. In particular, the source was undetected during a couple of time intervals on June 14 and was brighter during the second observation on June 15. A closer inspection of the time series on
June 15 revealed that the second X-ray flaring activity took place
$\lesssim$15\,min before a sharp peak in the radio emission (with a flux density of 266$\pm$27\,$\mu$Jy over a 10-min interval; Fig.\,\ref{fig:lc_zoom}).

The optical faintness of \src\ and the presence of a closeby extended scattered light feature in the OM images resulted in non-detections for most of the time span covered by the \xmm\ observations. Even in the time intervals where optical emission was detected in the second part of the second observation, the net count rates were too small to enable a detailed study of variability on timescales comparable with those of the X-ray mode switching.
While mild variability in the optical intensity can be seen between these images, no evidence for a significant flux enhancement was found around the epoch of the flares at X-ray and radio wavelengths, possibly due to the long integration time of the OM image ($\simeq$4\,ks; see the bottom panel in Fig.\,\ref{fig:lc}). However, recent multi-band, high-time resolution photometric observations of \src\ using ULTRACAM mounted on ESO-NTT revealed that the source indeed also shows optical flaring activity on similar timescales (Coti Zelati et al. in prep.). Hence, a higher time resolution is required to investigate possible connections (including correlations and lags) between the X-ray, radio and optical emissions during flares.

Overall, the radio and optical fluxes appear to be higher towards the end of our observing run on 2019 June 15, when the source spent most of the time in the X-ray high and flaring modes (Fig.\,\ref{fig:lc}). To assess the presence of possible correlated or anticorrelated variability between the radio and X-ray emissions, we adopted two different approaches on the data taken on June 15. Firstly, we performed the Spearman and Kendall $\tau$ rank correlation tests \citep{spearman04,kendall38} on the values of the radio fluxes and the X-ray count rates measured over strictly simultaneous 30\,min-long time intervals. We obtained two-sided $p$-values of $\simeq$0.40 and $\simeq$0.31, respectively, meaning that we cannot reject the null hypothesis that the two variables are uncorrelated at a high confidence level. Secondly, we measured the radio fluxes in the stacked images acquired during the periods of X-ray high mode as well as in the image acquired during the 50-min long period of X-ray low mode. We measured averaged flux densities of 59$\pm$5\,$\mu$Jy (12.0$\sigma$) in the former case and 50$\pm$12\,$\mu$Jy (4.2$\sigma$) in the latter case, that is, compatible with each other within the uncertainties\footnote{We note that these values are higher than the averaged flux density of $\simeq$33\,$\mu$Jy measured using the whole data sets. This discrepancy is due to the long-timescale variability in the radio emission, which yields a flux that is substantially higher on June 15 than on June 14 on average.}. Based on the above analyses, we conclude that no clear evidence for a significant correlated or anticorrelated variability is found between the radio and X-ray emissions on the sampled timescales.

\subsection{Searches for radio flares in the ATCA data}

A radio flare as bright as the one detected in our \meer\ data (see Fig.\,\ref{fig:lc}) should be easily at reach of the ATCA observations. Hence, we separated our ATCA observations into $\approx$30-min time intervals to search for any such radio flares. For all time intervals but one, no radio source was detected at the position of \src, with typical 3-$\sigma$ upper-limits of 80 $\mu$Jy. During one time interval on September 7 (between 02:54:00 and 03:36:00 UTC), we see a marginal detection consistent with the position of \src\ at a flux density of 94$\pm$25\,$\mu$Jy (3.8$\sigma$) at 7.25\,GHz. However, the limited instantaneous uv-coverage of ATCA resulted in a high image noise and an elongated beam, making it difficult to conclusively determine if the detection was real. Refining the time range, changing image weighting, fitting for a point source in the uv-plane, and limiting the image to just one of the observing bands (5.5 or 9\,GHz) did not enhance this possible detection. Hence, despite its detection fitting within the behaviour observed with \meer, we cannot convincingly determine whether this detection was real and we do not consider it further in our discussion, pending a higher significance detection at these frequencies in the future.

We note that the ATCA observations were taken at an epoch where \src\ was in
the same X-ray sub-luminous state observed a few months before during our \xmm\ and \meer\ observing campaign. Indeed, the background-subtracted X-ray spectrum acquired during the simultaneous \swift/XRT observations was well described by an absorbed PL model with $\Gamma=1.5\pm0.2$ (1$\sigma$ c.l.) and the average unabsorbed flux, $F_{\rm X}\simeq3.3\times10^{-12}$ \flux\ (0.3--10\,keV), was compatible with that measured in all previous X-ray observations during the accretion disk state (Fig.\,12 by \citetalias{cotizelati19}).

\begin{figure*}
\begin{center}
\includegraphics[width=.75\textwidth]{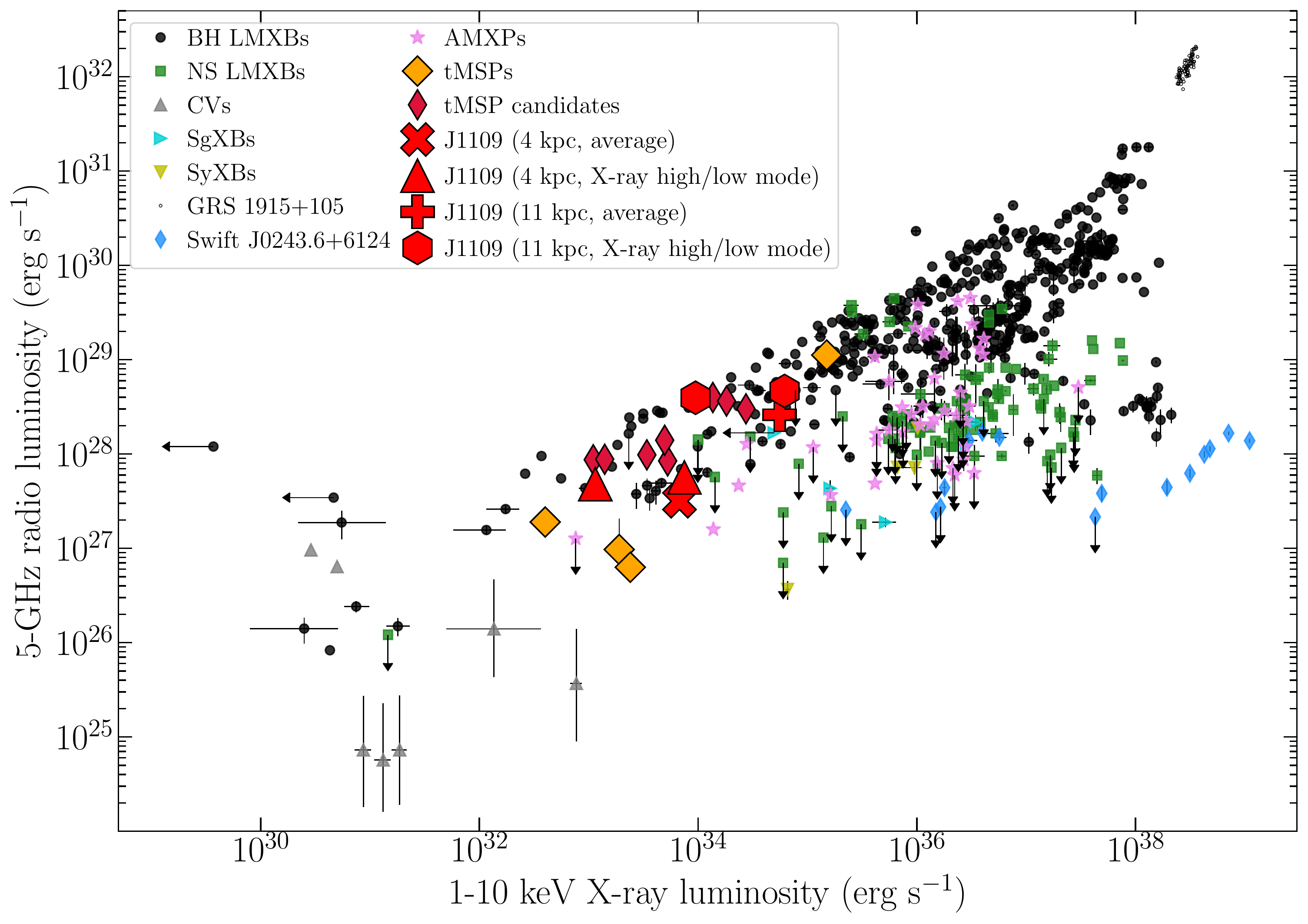}
\vspace{-0.1cm}
\caption{Radio and X-ray luminosities for different classes of binary systems harbouring different types of accreting compact objects, including low-mass X-ray binaries with BHs and NSs
accretors (BH LMXBs, NS LMXBs), cataclysmic variables (CVs), supergiant X-ray binaries (SgXBs), symbiotic X-ray binaries (SyXBs), the peculiar Be X-ray binary Swift\,J0243$+$6124, accreting millisecond X-ray pulsars
(AMXPs) and known and candidate tMSPs in the disk state. Values are
derived from simultaneous or quasi-simultaneous (within 1--2\,d) radio and X-ray observations, and are taken from \cite{bahramian18} with updates from 
\cite{bassi19,cotizelati19,gusinskaia19thesis,parikh19,bright20,gusinskaia20a,gusinskaia20b,hewitt20,li20,tremou20,williams20,xie20,vandene20,dehaas21,carotenuto21,motta21,paduano21,vandene21}.
Values for \src\ are plotted separately for the averaged emission and for the emission during the high and low X-ray modes.
Error bars on data points (when available) indicate 1$\sigma$ uncertainties, and are smaller than the marker size for most of the data points (including those corresponding to \src). Upper limits are marked using arrows.}
\label{fig:lxlr}
\end{center}
\end{figure*}

\subsection{The radio brightness of \src}
\label{lxlradio}

The simultaneous \xmm\ and \meer\ observations allow us to place \src\ on the X-ray versus radio luminosity plane for accreting compact objects. 
The zero-point corrected value for the source parallax has been recently updated to $\parallax = 449\pm637$\,$\mu$as in the {\it Gaia} EDR3 \citep{lindegren21,gaia2021}. For such a large uncertainty, the value for the inferred distance 
depends heavily on the assumed prior. In the hypothesis of an exponentially decreasing space density prior, this value converts to a median 
distance of 3.4\,kpc, with a 1$\sigma$ confidence interval of 1.9--6.6\,kpc. Assuming instead the photo-geometric prior introduced by \cite{bailerjones21}, 
we obtain a median distance of 11.1\,kpc and a 1$\sigma$ confidence interval of 8.9--13.9\,kpc\footnote{This prior accounts at once for the parallax, colour and 
apparent magnitude quoted in EDR3 by exploiting the fact that stars of a given colour have a restricted range of probable absolute magnitudes plus 
extinction. In the case of \src, the bulk of the optical emission is provided by the accretion disk rather than by the companion star, making the source bluer and brighter (see \citetalias{cotizelati19} for more details on optical spectroscopy of this system). Therefore, the distance derived for \src\ using this prior is likely to be subject to large uncertainties.}. We thus computed the X-ray and radio luminosities (both averaged and separately for the high and low X-ray modes) assuming conservatively two different values 
for the source distance, 4 and 11\,kpc. Lacking any information on the radio spectral shape, we also assumed a flat spectrum as observed in other tMSPs in the accretion disk state (see e.g. \citealt{deller15})
and in most accreting X-ray binaries in the hard state to rescale the radio luminosity at 5\,GHz (i.e. we assumed a spectral index of $\alpha=0$, where the flux density $F_{\nu}$ scales 
with the observing frequency $\nu$ as $F_{\nu}\propto\nu^{\alpha}$). The region occupied by \src\ on the $L_X$ -- $L_R$ 
plane (Fig.\,\ref{fig:lxlr}) is consistent with that of other tMSPs and black hole (BH) binaries in the hard state, providing further evidence 
that the tMSPs are generally brighter in the radio band than other accreting NSs for a given X-ray luminosity.

\section{Discussion}
\label{discussion}

We have presented the discovery of a variable radio counterpart to the tMSP candidate \srcfirst. Such a detection was achieved in the $L$-band 
using the \meer\ interferometer, while the source was in a X-ray sub-luminous state according to simultaneous observations with \xmm. The radio 
counterpart displayed variability on timescales of hours as well as a 10-min flare occurring just a few minutes after the detection of at least three consecutive intense flashes of X-rays within a time span of $\sim$2.5\,min.    
No evidence for emission at similar intensity levels was found at higher radio frequencies in multi-epoch observations performed with the ATCA 
(see also \citetalias{cotizelati19}). This is indicative either of a steep averaged radio spectrum ($\alpha \lesssim-0.5$) or, more likely, of variability across different epochs (assuming a flat averaged spectrum, $\alpha\simeq0$, at all epochs). In fact, changes by a factor of 
$\approx$2 and $\gtrsim$3 in the average flux density have been observed at distinct epochs in the tMSP \psr\ \citep{deller15} 
and in the candidate \rxsfirst\ (\rxs; \citealt{jaodand19thesis}), respectively.

No obvious correlated or anticorrelated variability pattern was detected between the radio and X-ray emissions in our \meer\ and \xmm\ data.
However, the radio variability properties, their connection with the emission observed at other wavelengths and the timescales involved may differ from source to source.
Substantial, rapid changes in the radio emission from a tMSP were first reported by \cite{ferrigno14}, who detected variations in the 
radio flux of the `swinging pulsar' \igrfirst\ (\igr) by a factor of $\sim$3 over one hour or so during its X-ray outburst in 2013. Moreover, these observations were quasi-simultaneous to X-ray observations and hinted at a possible anti-correlation between the radio and X-ray emissions. More recently, simultaneous X-ray and radio observations of the tMSP \psr\ in the X-ray sub-luminous state revealed a highly 
reproducible, anticorrelated variability pattern between the emissions in the two bands: the switches from the high to the low 
X-ray modes were always accompanied by enhancements in the radio brightness, with a duration that matches the low mode interval. These rebrightenings were detected using the Very Large Array at frequencies in the range 8--12\,GHz (i.e. higher than those probed for \src\ by \meer) and were associated with an evolution of the radio spectrum from inverted to relatively steep over the course of several minutes. Sporadic minutes-long radio flares were also observed \citep{bogdanov18}. A similar anti-correlation was also found in the candidate \rxs\ \citep{gusinskaia19thesis}, but not in the other candidate \fglfirst\ (\fgl; \citealt{li20}). We note, however, that \fgl\ showed a rather different phenomenology compared to the other tMSPs. Indeed, it was observed to spend most of the time in a flaring mode in the X-ray, UV and optical bands without showing the characteristic X-ray mode switching and displayed only modest variability in the radio emission on timescales as short as hours, possibly correlated with the X-ray flaring activity \citep{li20}. Very recently, the \fermi\ source 4FGL\,J0540.0$-$7552 (J0540) and the X-ray source NGC 6652B in the globular cluster NGC 6652 have been identified as tMSP candidates characterised by a multiband flaring activity that is remarkably similar to that observed in \fgl\ \citep{strader21,paduano21}. J0540 has not been observed at radio frequencies yet. Future coordinated multiband observations of this system including radio interferometers may allow the detection of radio emission and an assessment of a possible connection with the flaring emission detected at higher energies. NGC 6652B has been observed in the radio band at different epochs and displayed variations in the radio brightness by a factor of $\sim$2 on timescales of months. Additionally, simultaneous X-ray and radio observations detected numerous X-ray flares and a number of radio flares with typical durations of a few hundreds of seconds. However, no evidence for a significant correlation between these emissions was found.

The drop in the X-ray flux when switching from the high mode to the low mode is comparable in \src\ and the prototypical tMSP \psr\ (a factor of $\simeq$7; see Table\,\ref{tab:flux} and \citealt{papitto19} and references therein). However, the variability properties of the radio emission are seemingly different in the two systems. On the one hand, no obvious rebrightening in the radio band was observed in \src\ during the periods of X-ray low mode. On the other hand, enhancements in the radio brightness by a factor of $\sim$3 on average were detected in the case of \psr\ during all the periods of X-ray low mode, which lasted typically only a few minutes (i.e. they were much shorter than the 50 min-long episode of X-ray low mode observed from \src\ during our campaign; see \citealt{bogdanov18}). We note that our non-detection of any significant radio rebrightening during the periods of X-ray low mode in \src\ is not due to limitations in the instrumental detection sensitivity: the rms noise level attained by \meer\ during the long episode of X-ray low mode in the second observation was $\approx$12\,$\mu$Jy, hence a concurrent increase in the radio flux density of a similar extent as seen in \psr\ could have been detected at a confidence level of $\gtrsim$10$\sigma$. Overall, our analysis seems thus to suggest that the mechanism responsible for the radio emission at low frequencies in \src\ may not be directly connected with the process driving the X-ray mode switching. 

Assuming that the X-ray mode switching in the tMSPs truly operates close to the NS as suggested by recent studies \citep{papitto19,veledina19,campana19}, we may speculate that the bulk of the low-frequency emission of \src\ is provided by compact jet radiation and that possible additional ejection processes close to the NS (similar to those invoked for \psr) may still be at work but without producing a detectable increase in the radio brightness above the jet emission level. This scenario may well be possible if the ejected material remains optically thick to synchrotron self-absorption at the low frequencies probed by \meer\ ($\sim$1.3\,GHz) over the whole duration of the X-ray low modes.
In the future, deeper multi-frequency radio observations of \src\ with the Square Kilometre Array (SKA) coordinated with X-ray observations may allow testing for this scenario by characterising the broadband spectral evolution of the radio emission during the X-ray low modes. Additionally, a detection of possible turnovers in the spectral slope that lead to flux rebrightenings only above certain radio frequencies can also provide a valuable test for this scenario.

\subsection{The nature of the radio flare}

The nature of the enhanced radio emission observed in \src\ towards the end of 2019 June 15 is perhaps more uncertain. It may be associated with optically thin synchrotron radiation 
from discrete blobs of plasma ejected away from the compact object due to the formation of a transient accretion-driven jet or, alternatively, due to a mechanism ultimately connected with the presence of an active rotation-powered pulsar. The peculiar properties 
of the pulsed emission of \psr\ at optical and X-ray wavelengths, in fact, were recently interpreted in terms of synchrotron radiation from 
particles accelerated at a termination shock that forms due to the interaction between the particle wind ejected from the pulsar 
and plasma inflowing in the inner disk regions (a few hundreds of kilometers away from the pulsar, just outside the light cylinder; 
\citealt{papitto19,veledina19}). In this framework, X-ray emission during flares may be due to temporary enshrouding of the wind by the shock front 
(see also \citealt{campana19}), possibly due to a sudden increase in the thickness of the inner edge of the accretion disk \citep{veledina19}. In 
this phase, episodes of reconnection of magnetic field lines at the turbulent termination shock may launch outflows of 
plasmoids. For particularly strong, short-lived X-ray flares, a delayed flaring emission may be expected at radio frequencies (as observed in \src\ as well as in \psr; see Fig.\,\ref{fig:lc_zoom} and \citealt{bogdanov18}) as the plasma quickly expands outwards and becomes progressively more rarefied and transparent to synchrotron radiation even at low frequencies. Magnetic reconnection could also explain the increase in the optical flux observed around the same epoch of the X-ray flares and of the enhanced radio brightness (see Fig.\,\ref{fig:lc}).

We can tentatively estimate the physical parameters of the radio synchrotron-emitting plasma during flaring in the hypothesis that no energy is carried 
away by baryons and the plasma fulfills the equipartition condition, that is, the energy content in electrons balances that in the magnetic 
field. Assuming for simplicity a symmetric profile for the light curve of the flare (i.e. assuming a flare rise time of $\sim$5\,min, which is about half the duration of the time bin corresponding to the peak in the radio flux density; 
see Fig.\,\ref{fig:lc_zoom}), we can derive a minimum internal energy of $E_{{\rm min}}\approx10^{37}~d_4^{8/7}$\,erg and a magnetic field of 
$B\approx 0.3~d_4^{4/7}$\,G at equipartition (see \citealt{fender06}). We can also set a conservative lower limit on the linear size of the 
emitting region of $R>4\times10^{11}~d_4$\,cm by imposing that the brightness temperature $T_{\rm b}$ does not exceed the limit of 
$10^{12}$\,K, above which energy loss by synchrotron self-Compton radiation would become catastrophic \citep{longair11}. We obtain instead values of
$E_{{\rm min}}\approx10^{36}~d_4^{40/17}$\,erg, $B\approx0.3~d_4^{-4/17}$\,G and $R\approx2.6\times10^{12}~d_4^{16/17}$\,cm under
the additional hypothesis that the flare peak corresponds to an optical depth transition from thick to thin (see \citealt{fender19}). While the orbital parameters of \src\ are still unknown, it seems plausible that the two length scales estimated above could be larger than the orbital separation expected for this system\footnote{The orbital separation is of the order of a few 10$^{11}$\,cm in the case of \igr, the tMSP with the longest orbital period \citep{papitto13}.}, suggesting that the radio flaring emission is likely to be radiated from an extended cloud of plasma. We note, however, that the assumption of equipartition may well be an oversimplification. As a matter of fact, all pulsar wind nebulae detected at high energies are strongly particle-dominated and their magnetization is usually at the level of just a few percent (e.g. \citealt {Tanaka2010,Torres14}). Besides, caution should be exerted in overinterpreting the latter estimates in the absence of any measurement of the spectral shape and evolution of the flaring emission.

\section{Conclusions}
\label{conclusions}

We have presented the discovery of a variable radio counterpart to the tMSP candidate \src\ in multi-epoch coordinated X-ray and radio observations performed when the source was in a X-ray sub-luminous state. \src\ joins the small sample of known and candidate tMSPs for which relatively bright radio emission has been detected in the X-ray sub-luminous state. 

Future observations with the SKA may enable a more detailed investigation of variability in the emission of \src\ across a range of radio frequencies and on timescales compatible with those of the X-ray mode switching (see e.g. \citealt{braun19}). A conclusive assessment on the presence or absence of (anti)correlated variability between the radio and X-ray emissions in this system and other candidates will be critical in providing a consistent picture for the mechanisms of mass ejection and accretion in the tMSPs.

\begin{acknowledgements}
We thank N.~Schartel for approving Target of Opportunity observations with \xmm\ in the Director's Discretionary Time and the \xmm\ Science Operation Center
for scheduling and carrying out the observations. We thank S.~Goedhart and the staff at the South African Radio Astronomy Observatory (SARAO) for scheduling and carrying out the \meer\ observations. We thank the reviewer for providing helpful comments. \\
FCZ is supported by a Juan de la Cierva fellowship (IJC2019-042002-I). FCZ, DFT and NR are supported by the Spanish Grant PGC2018-095512-B-I00 and the Catalan Grant SGR2017-1383. FCZ and NR are also supported by the ERC Consolidator Grant `MAGNESIA' (nr. 817661).
DdM and AP acknowledge financial support from the Italian Space Agency (ASI) and National Institute for Astrophysics (INAF) under agreements ASI-INAF I/037/12/0 and ASI-INAF n.2017-14-H.0 (PI: Belloni), from INAF `Sostegno alla ricerca scientifica main streams dell'INAF' (PI: Belloni) Presidential Decree 43/2018, and from INAF `Towards the SKA and CTA era: discovery, localisation, and physics of transient sources' (PI: Giroletti) Presidential Decree 70/2016.
DAHB is supported by the National Research Foundation (NRF) of South Africa. TDR acknowledges financial contribution from the agreement ASI-INAF n.2017-14-H.0.
JL acknowledges the support from the National Natural Science Foundation of China via NSFC-11733009.
We acknowledge the International Space Science Institute (ISSI and ISSI -- Beijing), which funded and hosted the international teams `The disk-magnetosphere interaction around transitional millisecond pulsars' and `Understanding and unifying the gamma-rays emitting scenarios in high mass and low mass X-ray binaries'.
We thank partial support from the COST Action `PHAROS' (CA 16124).\\
The \xmm\ data are available at the European Space Agency (ESA) archive (\url{http://nxsa.esac.esa.int/nxsa-web}; obs. IDs: 0851180201, 0851180301). The \swift\ data are available at the \swift\ archive (\url{https://www.swift.ac.uk/swift_portal}; obs. ID: 00011274002). The \meer\ observations were obtained under the Open Time proposal SCI-20190418-DB-01 (PI: DAHB). The uncalibrated visibility data are available at the SARAO archive (\url{https:
//archive.sarao.ac.za}). The Australia Telescope Compact Array (ATCA) observations were obtained under project code C3007 (PI: FCZ). The uncalibrated visibility data are available at the Australia Telescope National Facility (ATNF) archive (\url{https://atoa.atnf.csiro.au}). The calibrated \meer\ data and images that support the findings of this study, the data for the light curves and the values for the data points shown in Fig.\,\ref{fig:lxlr} of this paper are available from the corresponding author upon reasonable request.\\
\xmm\ is an ESA science mission with instruments and contributions directly funded by ESA Member States and NASA. The {\em Neil Gehrels Swift Observatory} is a NASA/UK/ASI mission. The \meer\ telescope is operated by SARAO, which is a facility of the NRF, an agency of the Department of Science and Innovation. 
The ATCA is part of the ATNF, which is funded by the Australian Government for 
operation as a National Facility managed by the Commonwealth Scientific and Industrial Research Organisation. We acknowledge the Gomeroi people as the traditional owners of the ATCA observatory site.\\
This research has made use of the following software:
APLPY v.2.0.3 \citep{robitaille12}, Astropy v.4.3.1 \citep{astropy:2013, astropy:2018}, CASA v.5.6.0 \citep{mcmullin07}, cleanmask v.1.3.1 
(\url{https://github.com/SpheMakh/cleanmask}), Cubical v.1.5.11 \citep{kenyon18}, DDFacet v.0.6.0 \citep{tasse18}, HEASOFT v.6.29c, katdal v.0.18 (\url{https://github.com/ska-sa/katdal}), MATPLOTLIB v.3.4.3 \citep{hunter07}, NUMPY v.1.21.2 \citep{harris20}, pyBDSF v.1.9.1 \citep{mohan15}, 
SAOImageDS9 v.8.2.1 \citep{joye03}, SAS v.19 \citep{gabriel04}, SCIPY v.1.7.1 \citep{scipy20}, Tigger v.1.6.0 (\url{https://github.com/ska-sa/tigger}), XRONOS v.5.21 \citep{stella92}, XSPEC v.12.12.0 \citep{arnaud96}.
\end{acknowledgements}

\vspace{-0.5cm}

\bibliographystyle{aa} 
\bibliography{biblio}

\begin{thebibliography}{81}
\expandafter\ifx\csname natexlab\endcsname\relax\def\natexlab#1{#1}\fi

\bibitem[{{Ambrosino} {et~al.}(2017){Ambrosino}, {Papitto}, {Stella}, {Meddi},
  {Cretaro}, {Burderi}, {Di Salvo}, {Israel}, {Ghedina}, {Di Fabrizio}, \&
  {Riverol}}]{ambrosino17}
{Ambrosino}, F., {Papitto}, A., {Stella}, L., {et~al.} 2017,
  \href{http://dx.doi.org/10.1038/s41550-017-0266-2}{\color{magenta}Nat.
  Astron.}, \href{https://ui.adsabs.harvard.edu/abs/2017NatAs...1..854A}{1,
  854}

\bibitem[{{Archibald} {et~al.}(2009){Archibald}, {Stairs}, {Ransom}, {Kaspi},
  {Kondratiev}, {Lorimer}, {McLaughlin}, {Boyles}, {Hessels}, \&
  {Lynch}}]{archibald09}
{Archibald}, A.~M., {Stairs}, I.~H., {Ransom}, S.~M., {et~al.} 2009,
  \href{http://dx.doi.org/10.1126/science.1172740}{\color{magenta}Science},
  \href{https://ui.adsabs.harvard.edu/abs/2009Sci...324.1411A}{324, 1411}

\bibitem[{{Arnaud}(1996)}]{arnaud96}
{Arnaud}, K.~A. 1996, in Astronomical Society of the Pacific Conference Series,
  Vol. 101, Astronomical Data Analysis Software and Systems V, ed. G.~H.
  {Jacoby} \& J.~{Barnes},
  \href{http://adsabs.harvard.edu/abs/1996ASPC..101...17A}{17}

\bibitem[{{Astropy Collaboration} {et~al.}(2018){Astropy Collaboration},
  {Price-Whelan}, {Sip\H{o}cz}, {G{"u}nther}, {Lim}, {Crawford}, {Conseil},
  {Shupe}, {Craig}, {Dencheva}, {Ginsburg}, {Vand erPlas}, {Bradley},
  {P{'e}rez-Su{'a}rez}, {de Val-Borro}, {Aldcroft}, {Cruz}, {Robitaille},
  {Tollerud}, {Ardelean}, {Babej}, {Bach}, {Bachetti}, {Bakanov}, {Bamford},
  {Barentsen}, {Barmby}, {Baumbach}, {Berry}, {Biscani}, {Boquien}, {Bostroem},
  {Bouma}, {Brammer}, {Bray}, {Breytenbach}, {Buddelmeijer}, {Burke},
  {Calderone}, {Cano Rodr{'i}guez}, {Cara}, {Cardoso}, {Cheedella}, {Copin},
  {Corrales}, {Crichton}, {D'Avella}, {Deil}, {Depagne}, {Dietrich}, {Donath},
  {Droettboom}, {Earl}, {Erben}, {Fabbro}, {Ferreira}, {Finethy}, {Fox},
  {Garrison}, {Gibbons}, {Goldstein}, {Gommers}, {Greco}, {Greenfield},
  {Groener}, {Grollier}, {Hagen}, {Hirst}, {Homeier}, {Horton}, {Hosseinzadeh},
  {Hu}, {Hunkeler}, {Ivezi{'c}}, {Jain}, {Jenness}, {Kanarek}, {Kendrew},
  {Kern}, {Kerzendorf}, {Khvalko}, {King}, {Kirkby}, {Kulkarni}, {Kumar},
  {Lee}, {Lenz}, {Littlefair}, {Ma}, {Macleod}, {Mastropietro}, {McCully},
  {Montagnac}, {Morris}, {Mueller}, {Mumford}, {Muna}, {Murphy}, {Nelson},
  {Nguyen}, {Ninan}, {N{"o}the}, {Ogaz}, {Oh}, {Parejko}, {Parley}, {Pascual},
  {Patil}, {Patil}, {Plunkett}, {Prochaska}, {Rastogi}, {Reddy Janga},
  {Sabater}, {Sakurikar}, {Seifert}, {Sherbert}, {Sherwood-Taylor}, {Shih},
  {Sick}, {Silbiger}, {Singanamalla}, {Singer}, {Sladen}, {Sooley},
  {Sornarajah}, {Streicher}, {Teuben}, {Thomas}, {Tremblay}, {Turner},
  {Terr{'o}n}, {van Kerkwijk}, {de la Vega}, {Watkins}, {Weaver}, {Whitmore},
  {Woillez}, {Zabalza}, \& {Astropy Contributors}}]{astropy:2018}
{Astropy Collaboration}, {Price-Whelan}, A.~M., {Sip\H{o}cz}, B.~M., {et~al.}
  2018, \href{http://dx.doi.org/10.3847/1538-3881/aabc4f}{\color{magenta}AJ},
  \href{https://ui.adsabs.harvard.edu/abs/2018AJ....156..123A}{156, 123}

\bibitem[{{Astropy Collaboration} {et~al.}(2013){Astropy Collaboration},
  {Robitaille}, {Tollerud}, {Greenfield}, {Droettboom}, {Bray}, {Aldcroft},
  {Davis}, {Ginsburg}, {Price-Whelan}, {Kerzendorf}, {Conley}, {Crighton},
  {Barbary}, {Muna}, {Ferguson}, {Grollier}, {Parikh}, {Nair}, {Unther},
  {Deil}, {Woillez}, {Conseil}, {Kramer}, {Turner}, {Singer}, {Fox}, {Weaver},
  {Zabalza}, {Edwards}, {Azalee Bostroem}, {Burke}, {Casey}, {Crawford},
  {Dencheva}, {Ely}, {Jenness}, {Labrie}, {Lim}, {Pierfederici}, {Pontzen},
  {Ptak}, {Refsdal}, {Servillat}, \& {Streicher}}]{astropy:2013}
{Astropy Collaboration}, {Robitaille}, T.~P., {Tollerud}, E.~J., {et~al.} 2013,
  \href{http://dx.doi.org/10.1051/0004-6361/201322068}{\color{magenta}\aap},
  \href{http://adsabs.harvard.edu/abs/2013A%26A...558A..33A}{558, A33}

\bibitem[{{Baglio} {et~al.}(2019){Baglio}, {Vincentelli}, {Campana}, {Coti
  Zelati}, {D'Avanzo}, {Burderi}, {Casella}, {Papitto}, \&
  {Russell}}]{baglio19}
{Baglio}, M.~C., {Vincentelli}, F., {Campana}, S., {et~al.} 2019,
  \href{http://dx.doi.org/10.1051/0004-6361/201936008}{\color{magenta}\aap},
  \href{https://ui.adsabs.harvard.edu/abs/2019A&A...631A.104B}{631, A104}

\bibitem[{Bahramian {et~al.}(2018)Bahramian, Miller-Jones, Strader, Tetarenko,
  Plotkin, Rushton, Tudor, Motta, \& Shishkovsky}]{bahramian18}
Bahramian, A., Miller-Jones, J., Strader, J., {et~al.} 2018, {Radio/X-ray
  correlation database for X-ray binaries}

\bibitem[{{Bailer-Jones} {et~al.}(2021){Bailer-Jones}, {Rybizki}, {Fouesneau},
  {Demleitner}, \& {Andrae}}]{bailerjones21}
{Bailer-Jones}, C.~A.~L., {Rybizki}, J., {Fouesneau}, M., {Demleitner}, M., \&
  {Andrae}, R. 2021,
  \href{http://dx.doi.org/10.3847/1538-3881/abd806}{\color{magenta}\aj},
  \href{https://ui.adsabs.harvard.edu/abs/2021AJ....161..147B}{161, 147}

\bibitem[{{Bassa} {et~al.}(2014){Bassa}, {Patruno}, {Hessels}, {Keane},
  {Monard}, {Mahony}, {Bogdanov}, {Corbel}, {Edwards}, \&
  {Archibald}}]{bassa14}
{Bassa}, C.~G., {Patruno}, A., {Hessels}, J.~W.~T., {et~al.} 2014,
  \href{http://dx.doi.org/10.1093/mnras/stu708}{\color{magenta}\mnras},
  \href{https://ui.adsabs.harvard.edu/abs/2014MNRAS.441.1825B}{441, 1825}

\bibitem[{{Bassi} {et~al.}(2019){Bassi}, {Del Santo}, {D'A{\`i}}, {Motta},
  {Malzac}, {Segreto}, {Miller-Jones}, {Atri}, {Plotkin}, {Belloni}, {Mineo},
  \& {Tzioumis}}]{bassi19}
{Bassi}, T., {Del Santo}, M., {D'A{\`i}}, A., {et~al.} 2019,
  \href{http://dx.doi.org/10.1093/mnras/sty2739}{\color{magenta}MNRAS},
  \href{https://ui.adsabs.harvard.edu/abs/2019MNRAS.482.1587B}{482, 1587}

\bibitem[{{Blandford} \& {K{\"o}nigl}(1979)}]{blandford79}
{Blandford}, R.~D. \& {K{\"o}nigl}, A. 1979,
  \href{http://dx.doi.org/10.1086/157262}{\color{magenta}\apj},
  \href{https://ui.adsabs.harvard.edu/abs/1979ApJ...232...34B}{232, 34}

\bibitem[{{Blandford} \& {Payne}(1982)}]{blandford82}
{Blandford}, R.~D. \& {Payne}, D.~G. 1982,
  \href{http://dx.doi.org/10.1093/mnras/199.4.883}{\color{magenta}\mnras},
  \href{https://ui.adsabs.harvard.edu/abs/1982MNRAS.199..883B}{199, 883}

\bibitem[{Bogdanov {et~al.}(2018)Bogdanov, Deller, Miller-Jones, Archibald,
  Hessels, Jaodand, Patruno, Bassa, \& D'Angelo}]{bogdanov18}
Bogdanov, S., Deller, A.~T., Miller-Jones, J. C.~A., {et~al.} 2018,
  \href{http://dx.doi.org/10.3847/1538-4357/aaaeb9}{\color{magenta}ApJ}, 856,
  856

\bibitem[{{Braun} {et~al.}(2019){Braun}, {Bonaldi}, {Bourke}, {Keane}, \&
  {Wagg}}]{braun19}
{Braun}, R., {Bonaldi}, A., {Bourke}, T., {Keane}, E., \& {Wagg}, J. 2019,
  \href{https://ui.adsabs.harvard.edu/abs/2019arXiv191212699B}{arXiv e-prints,
  arXiv:1912.12699}

\bibitem[{{Briggs}(1995)}]{briggs95}
{Briggs}, D.~S. 1995, in American Astronomical Society Meeting Abstracts, Vol.
  187, American Astronomical Society Meeting Abstracts,
  \href{https://ui.adsabs.harvard.edu/abs/1995AAS...18711202B}{112.02}

\bibitem[{{Bright} {et~al.}(2020){Bright}, {Fender}, {Motta}, {Williams},
  {Moldon}, {Plotkin}, {Miller-Jones}, {Heywood}, {Tremou}, {Beswick},
  {Sivakoff}, {Corbel}, {Buckley}, {Homan}, {Gallo}, {Tetarenko}, {Russell},
  {Green}, {Titterington}, {Woudt}, {Armstrong}, {Groot}, {Horesh}, {van der
  Horst}, {K{\"o}rding}, {McBride}, {Rowlinson}, \& {Wijers}}]{bright20}
{Bright}, J.~S., {Fender}, R.~P., {Motta}, S.~E., {et~al.} 2020,
  \href{http://dx.doi.org/10.1038/s41550-020-1023-5}{\color{magenta}Nat.
  Astron.}, \href{https://ui.adsabs.harvard.edu/abs/2020NatAs...4..697B}{4,
  697}

\bibitem[{{Burrows} {et~al.}(2005){Burrows}, {Hill}, {Nousek}, {Kennea},
  {Wells}, {Osborne}, {Abbey}, {Beardmore}, {Mukerjee}, {Short}, {Chincarini},
  {Campana}, {Citterio}, {Moretti}, {Pagani}, {Tagliaferri}, {Giommi},
  {Capalbi}, {Tamburelli}, {Angelini}, {Cusumano}, {Br{\"a}uninger}, {Burkert},
  \& {Hartner}}]{burrows05}
{Burrows}, D.~N., {Hill}, J.~E., {Nousek}, J.~A., {et~al.} 2005,
  \href{http://dx.doi.org/10.1007/s11214-005-5097-2}{\color{magenta}\ssr},
  \href{https://ui.adsabs.harvard.edu/abs/2005SSRv..120..165B}{120, 165}

\bibitem[{{Camilo} {et~al.}(2018){Camilo}, {Scholz}, {Serylak}, {Buchner},
  {Merryfield}, {Kaspi}, {Archibald}, {Bailes}, {Jameson}, \& {van
  Straten}}]{camilo18}
{Camilo}, F., {Scholz}, P., {Serylak}, M., {et~al.} 2018,
  \href{http://dx.doi.org/10.3847/1538-4357/aab35a}{\color{magenta}\apj},
  \href{https://ui.adsabs.harvard.edu/abs/2018ApJ...856..180C}{856, 180}

\bibitem[{{Campana} {et~al.}(2016){Campana}, {Coti Zelati}, {Papitto}, {Rea},
  {Torres}, {Baglio}, \& {D'Avanzo}}]{campana16}
{Campana}, S., {Coti Zelati}, F., {Papitto}, A., {et~al.} 2016,
  \href{http://dx.doi.org/10.1051/0004-6361/201629035}{\color{magenta}\aap},
  \href{https://ui.adsabs.harvard.edu/abs/2016A&A...594A..31C}{594, A31}

\bibitem[{{Campana} \& {Di Salvo}(2018)}]{campana18}
{Campana}, S. \& {Di Salvo}, T. 2018, in Astrophysics and Space Science
  Library, Vol. 457, Astrophysics and Space Science Library, ed. L.~{Rezzolla},
  P.~{Pizzochero}, D.~I. {Jones}, N.~{Rea}, \& I.~{Vida{\~n}a},
  \href{https://ui.adsabs.harvard.edu/abs/2018ASSL..457..149C}{149}

\bibitem[{{Campana} {et~al.}(2019){Campana}, {Miraval Zanon}, {Coti Zelati},
  {Torres}, {Baglio}, \& {Papitto}}]{campana19}
{Campana}, S., {Miraval Zanon}, A., {Coti Zelati}, F., {et~al.} 2019,
  \href{http://dx.doi.org/10.1051/0004-6361/201936312}{\color{magenta}\aap},
  \href{https://ui.adsabs.harvard.edu/abs/2019A&A...629L...8C}{629, L8}

\bibitem[{{Carotenuto} {et~al.}(2021){Carotenuto}, {Corbel}, {Tremou},
  {Russell}, {Tzioumis}, {Fender}, {Woudt}, {Motta}, {Miller-Jones},
  {Tetarenko}, \& {Sivakoff}}]{carotenuto21}
{Carotenuto}, F., {Corbel}, S., {Tremou}, E., {et~al.} 2021,
  \href{http://dx.doi.org/10.1093/mnrasl/slab049}{\color{magenta}\mnras},
  \href{https://ui.adsabs.harvard.edu/abs/2021MNRAS.505L..58C}{505, L58}

\bibitem[{{Coti Zelati} {et~al.}(2018){Coti Zelati}, {Campana}, {Braito},
  {Baglio}, {D'Avanzo}, {Rea}, \& {Torres}}]{cotizelati18}
{Coti Zelati}, F., {Campana}, S., {Braito}, V., {et~al.} 2018,
  \href{http://dx.doi.org/10.1051/0004-6361/201732244}{\color{magenta}\aap},
  \href{https://ui.adsabs.harvard.edu/abs/2018A&A...611A..14C}{611, A14}

\bibitem[{{Coti Zelati} {et~al.}(2019){Coti Zelati}, {Papitto}, {de Martino},
  {Buckley}, {Odendaal}, {Li}, {Russell}, {Torres}, {Mazzola}, \&
  {Bozzo}}]{cotizelati19}
{Coti Zelati}, F., {Papitto}, A., {de Martino}, D., {et~al.} 2019,
  \href{http://dx.doi.org/10.1051/0004-6361/201834835}{\color{magenta}\aap},
  \href{https://ui.adsabs.harvard.edu/abs/2019A&A...622A.211C}{622, A211}

\bibitem[{{de Haas} {et~al.}(2021){de Haas}, {Russell}, {Degenaar}, {Markoff},
  {Tetarenko}, {Tetarenko}, {van den Eijnden}, {Miller-Jones}, {Parikh},
  {Plotkin}, \& {Sivakoff}}]{dehaas21}
{de Haas}, S.~E.~M., {Russell}, T.~D., {Degenaar}, N., {et~al.} 2021,
  \href{http://dx.doi.org/10.1093/mnras/staa3853}{\color{magenta}\mnras},
  \href{https://ui.adsabs.harvard.edu/abs/2021MNRAS.502..521D}{502, 521}

\bibitem[{{de Martino} {et~al.}(2020){de Martino}, {Papitto}, {Burgay},
  {Possenti}, {Coti Zelati}, {Rea}, {Torres}, \& {Belloni}}]{demartino20}
{de Martino}, D., {Papitto}, A., {Burgay}, M., {et~al.} 2020,
  \href{http://dx.doi.org/10.1093/mnras/staa164}{\color{magenta}\mnras},
  \href{https://ui.adsabs.harvard.edu/abs/2020MNRAS.492.5607D}{492, 5607}

\bibitem[{{Deller} {et~al.}(2015){Deller}, {Moldon}, {Miller-Jones}, {Patruno},
  {Hessels}, {Archibald}, {Paragi}, {Heald}, \& {Vilchez}}]{deller15}
{Deller}, A.~T., {Moldon}, J., {Miller-Jones}, J.~C.~A., {et~al.} 2015,
  \href{http://dx.doi.org/10.1088/0004-637X/809/1/13}{\color{magenta}\apj},
  \href{https://ui.adsabs.harvard.edu/abs/2015ApJ...809...13D}{809, 13}

\bibitem[{{Fender}(2006)}]{fender06}
{Fender}, R. 2006, {Jets from X-ray binaries}, Vol.~39, 381--419

\bibitem[{{Fender} \& {Bright}(2019)}]{fender19}
{Fender}, R. \& {Bright}, J. 2019,
  \href{http://dx.doi.org/10.1093/mnras/stz2000}{\color{magenta}\mnras},
  \href{https://ui.adsabs.harvard.edu/abs/2019MNRAS.489.4836F}{489, 4836}

\bibitem[{{Ferrigno} {et~al.}(2014){Ferrigno}, {Bozzo}, {Papitto}, {Rea},
  {Pavan}, {Campana}, {Wieringa}, {Filipovi{\'c}}, {Falanga}, \&
  {Stella}}]{ferrigno14}
{Ferrigno}, C., {Bozzo}, E., {Papitto}, A., {et~al.} 2014,
  \href{http://dx.doi.org/10.1051/0004-6361/201322904}{\color{magenta}\aap},
  \href{https://ui.adsabs.harvard.edu/abs/2014A&A...567A..77F}{567, A77}

\bibitem[{{Gabriel} {et~al.}(2004){Gabriel}, {Denby}, {Fyfe}, {Hoar}, {Ibarra},
  {Ojero}, {Osborne}, {Saxton}, {Lammers}, \& {Vacanti}}]{gabriel04}
{Gabriel}, C., {Denby}, M., {Fyfe}, D.~J., {et~al.} 2004, in Astronomical
  Society of the Pacific Conference Series, Vol. 314, Astronomical Data
  Analysis Software and Systems (ADASS) XIII, ed. F.~{Ochsenbein}, M.~G.
  {Allen}, \& D.~{Egret},
  \href{https://ui.adsabs.harvard.edu/abs/2004ASPC..314..759G}{759}

\bibitem[{{Gaia Collaboration} {et~al.}(2021){Gaia Collaboration}, {Brown},
  {Vallenari}, {Prusti}, {de Bruijne}, {Babusiaux}, {Biermann}, {Creevey},
  {Evans}, {Eyer}, {Hutton}, {Jansen}, {Jordi}, {Klioner}, {Lammers},
  {Lindegren}, {Luri}, {Mignard}, {Panem}, {Pourbaix}, {Randich}, {Sartoretti},
  {Soubiran}, {Walton}, {Arenou}, {Bailer-Jones}, {Bastian}, {Cropper},
  {Drimmel}, {Katz}, {Lattanzi}, {van Leeuwen}, {Bakker}, {Cacciari},
  {Casta{\~n}eda}, {De Angeli}, {Ducourant}, {Fabricius}, {Fouesneau},
  {Fr{\'e}mat}, {Guerra}, {Guerrier}, {Guiraud}, {Jean-Antoine Piccolo},
  {Masana}, {Messineo}, {Mowlavi}, {Nicolas}, {Nienartowicz}, {Pailler},
  {Panuzzo}, {Riclet}, {Roux}, {Seabroke}, {Sordo}, {Tanga}, {Th{\'e}venin},
  {Gracia-Abril}, {Portell}, {Teyssier}, {Altmann}, {Andrae}, {Bellas-Velidis},
  {Benson}, {Berthier}, {Blomme}, {Brugaletta}, {Burgess}, {Busso}, {Carry},
  {Cellino}, {Cheek}, {Clementini}, {Damerdji}, {Davidson}, {Delchambre},
  {Dell'Oro}, {Fern{\'a}ndez-Hern{\'a}ndez}, {Galluccio}, {Garc{\'\i}a-Lario},
  {Garcia-Reinaldos}, {Gonz{\'a}lez-N{\'u}{\~n}ez}, {Gosset}, {Haigron},
  {Halbwachs}, {Hambly}, {Harrison}, {Hatzidimitriou}, {Heiter},
  {Hern{\'a}ndez}, {Hestroffer}, {Hodgkin}, {Holl}, {Jan{\ss}en}, {Jevardat de
  Fombelle}, {Jordan}, {Krone-Martins}, {Lanzafame}, {L{\"o}ffler}, {Lorca},
  {Manteiga}, {Marchal}, {Marrese}, {Moitinho}, {Mora}, {Muinonen}, {Osborne},
  {Pancino}, {Pauwels}, {Petit}, {Recio-Blanco}, {Richards}, {Riello},
  {Rimoldini}, {Robin}, {Roegiers}, {Rybizki}, {Sarro}, {Siopis}, {Smith},
  {Sozzetti}, {Ulla}, {Utrilla}, {van Leeuwen}, {van Reeven}, {Abbas}, {Abreu
  Aramburu}, {Accart}, {Aerts}, {Aguado}, {Ajaj}, {Altavilla}, {{\'A}lvarez},
  {{\'A}lvarez Cid-Fuentes}, {Alves}, {Anderson}, {Anglada Varela}, {Antoja},
  {Audard}, {Baines}, {Baker}, {Balaguer-N{\'u}{\~n}ez}, {Balbinot}, {Balog},
  {Barache}, {Barbato}, {Barros}, {Barstow}, {Bartolom{\'e}}, {Bassilana},
  {Bauchet}, {Baudesson-Stella}, {Becciani}, {Bellazzini}, {Bernet}, {Bertone},
  {Bianchi}, {Blanco-Cuaresma}, {Boch}, {Bombrun}, {Bossini}, {Bouquillon},
  {Bragaglia}, {Bramante}, {Breedt}, {Bressan}, {Brouillet}, {Bucciarelli},
  {Burlacu}, {Busonero}, {Butkevich}, {Buzzi}, {Caffau}, {Cancelliere},
  {C{\'a}novas}, {Cantat-Gaudin}, {Carballo}, {Carlucci}, {Carnerero},
  {Carrasco}, {Casamiquela}, {Castellani}, {Castro-Ginard}, {Castro Sampol},
  {Chaoul}, {Charlot}, {Chemin}, {Chiavassa}, {Cioni}, {Comoretto}, {Cooper},
  {Cornez}, {Cowell}, {Crifo}, {Crosta}, {Crowley}, {Dafonte}, {Dapergolas},
  {David}, {David}, {de Laverny}, {De Luise}, {De March}, {De Ridder}, {de
  Souza}, {de Teodoro}, {de Torres}, {del Peloso}, {del Pozo}, {Delbo},
  {Delgado}, {Delgado}, {Delisle}, {Di Matteo}, {Diakite}, {Diener},
  {Distefano}, {Dolding}, {Eappachen}, {Edvardsson}, {Enke}, {Esquej}, {Fabre},
  {Fabrizio}, {Faigler}, {Fedorets}, {Fernique}, {Fienga}, {Figueras},
  {Fouron}, {Fragkoudi}, {Fraile}, {Franke}, {Gai}, {Garabato},
  {Garcia-Gutierrez}, {Garc{\'\i}a-Torres}, {Garofalo}, {Gavras}, {Gerlach},
  {Geyer}, {Giacobbe}, {Gilmore}, {Girona}, {Giuffrida}, {Gomel}, {Gomez},
  {Gonzalez-Santamaria}, {Gonz{\'a}lez-Vidal}, {Granvik},
  {Guti{\'e}rrez-S{\'a}nchez}, {Guy}, {Hauser}, {Haywood}, {Helmi}, {Hidalgo},
  {Hilger}, {H{\l}adczuk}, {Hobbs}, {Holland}, {Huckle}, {Jasniewicz},
  {Jonker}, {Juaristi Campillo}, {Julbe}, {Karbevska}, {Kervella}, {Khanna},
  {Kochoska}, {Kontizas}, {Kordopatis}, {Korn}, {Kostrzewa-Rutkowska},
  {Kruszy{\'n}ska}, {Lambert}, {Lanza}, {Lasne}, {Le Campion}, {Le Fustec},
  {Lebreton}, {Lebzelter}, {Leccia}, {Leclerc}, {Lecoeur-Taibi}, {Liao},
  {Licata}, {Lindstr{\o}m}, {Lister}, {Livanou}, {Lobel}, {Madrero Pardo},
  {Managau}, {Mann}, {Marchant}, {Marconi}, {Marcos Santos}, {Marinoni},
  {Marocco}, {Marshall}, {Martin Polo}, {Mart{\'\i}n-Fleitas}, {Masip},
  {Massari}, {Mastrobuono-Battisti}, {Mazeh}, {McMillan}, {Messina},
  {Michalik}, {Millar}, {Mints}, {Molina}, {Molinaro}, {Moln{\'a}r},
  {Montegriffo}, {Mor}, {Morbidelli}, {Morel}, {Morris}, {Mulone}, {Munoz},
  {Muraveva}, {Murphy}, {Musella}, {Noval}, {Ord{\'e}novic}, {Orr{\`u}},
  {Osinde}, {Pagani}, {Pagano}, {Palaversa}, {Palicio}, {Panahi}, {Pawlak},
  {Pe{\~n}alosa Esteller}, {Penttil{\"a}}, {Piersimoni}, {Pineau}, {Plachy},
  {Plum}, {Poggio}, {Poretti}, {Poujoulet}, {Pr{\v{s}}a}, {Pulone}, {Racero},
  {Ragaini}, {Rainer}, {Raiteri}, {Rambaux}, {Ramos}, {Ramos-Lerate}, {Re
  Fiorentin}, {Regibo}, {Reyl{\'e}}, {Ripepi}, {Riva}, {Rixon}, {Robichon},
  {Robin}, {Roelens}, {Rohrbasser}, {Romero-G{\'o}mez}, {Rowell}, {Royer},
  {Rybicki}, {Sadowski}, {Sagrist{\`a} Sell{\'e}s}, {Sahlmann}, {Salgado},
  {Salguero}, {Samaras}, {Sanchez Gimenez}, {Sanna}, {Santove{\~n}a},
  {Sarasso}, {Schultheis}, {Sciacca}, {Segol}, {Segovia}, {S{\'e}gransan},
  {Semeux}, {Shahaf}, {Siddiqui}, {Siebert}, {Siltala}, {Slezak}, {Smart},
  {Solano}, {Solitro}, {Souami}, {Souchay}, {Spagna}, {Spoto}, {Steele},
  {Steidelm{\"u}ller}, {Stephenson}, {S{\"u}veges}, {Szabados}, {Szegedi-Elek},
  {Taris}, {Tauran}, {Taylor}, {Teixeira}, {Thuillot}, {Tonello}, {Torra},
  {Torra}, {Turon}, {Unger}, {Vaillant}, {van Dillen}, {Vanel}, {Vecchiato},
  {Viala}, {Vicente}, {Voutsinas}, {Weiler}, {Wevers}, {Wyrzykowski}, {Yoldas},
  {Yvard}, {Zhao}, {Zorec}, {Zucker}, {Zurbach}, \& {Zwitter}}]{gaia2021}
{Gaia Collaboration}, {Brown}, A.~G.~A., {Vallenari}, A., {et~al.} 2021,
  \href{http://dx.doi.org/10.1051/0004-6361/202039657}{\color{magenta}\aap},
  \href{https://ui.adsabs.harvard.edu/abs/2021A&A...649A...1G}{649, A1}

\bibitem[{{Gallo} {et~al.}(2018){Gallo}, {Degenaar}, \& {van den
  Eijnden}}]{gallo18}
{Gallo}, E., {Degenaar}, N., \& {van den Eijnden}, J. 2018,
  \href{http://dx.doi.org/10.1093/mnrasl/sly083}{\color{magenta}\mnras},
  \href{https://ui.adsabs.harvard.edu/abs/2018MNRAS.478L.132G}{478, L132}

\bibitem[{Gusinskaia(2019)}]{gusinskaia19thesis}
Gusinskaia, N. 2019, {Probing gravity and accretion using neutron stars}, PhD
  thesis, U. Amsterdam

\bibitem[{{Gusinskaia} {et~al.}(2020{\natexlab{a}}){Gusinskaia}, {Hessels},
  {Degenaar}, {Deller}, {Miller-Jones}, {Archibald}, {Heinke}, {Mold{\'o}n},
  {Patruno}, {Tomsick}, \& {Wijnands}}]{gusinskaia20a}
{Gusinskaia}, N.~V., {Hessels}, J.~W.~T., {Degenaar}, N., {et~al.}
  2020{\natexlab{a}},
  \href{http://dx.doi.org/10.1093/mnras/stz3420}{\color{magenta}\mnras},
  \href{https://ui.adsabs.harvard.edu/abs/2020MNRAS.492.2858G}{492, 2858}

\bibitem[{{Gusinskaia} {et~al.}(2020{\natexlab{b}}){Gusinskaia}, {Russell},
  {Hessels}, {Bogdanov}, {Degenaar}, {Deller}, {van den Eijnden}, {Jaodand},
  {Miller-Jones}, \& {Wijnands}}]{gusinskaia20b}
{Gusinskaia}, N.~V., {Russell}, T.~D., {Hessels}, J.~W.~T., {et~al.}
  2020{\natexlab{b}},
  \href{http://dx.doi.org/10.1093/mnras/stz3460}{\color{magenta}\mnras},
  \href{https://ui.adsabs.harvard.edu/abs/2020MNRAS.492.1091G}{492, 1091}

\bibitem[{{Harding}(2021)}]{harding21}
{Harding}, A.~K. 2021,
  \href{https://ui.adsabs.harvard.edu/abs/2021arXiv210105751H}{arXiv e-prints,
  arXiv:2101.05751}

\bibitem[{Harris {et~al.}(2020)Harris, Millman, van~der Walt, Gommers,
  Virtanen, Cournapeau, Wieser, Taylor, Berg, Smith, Kern, Picus, Hoyer, van
  Kerkwijk, Brett, Haldane, del R{'{\i}}o, Wiebe, Peterson,
  G{'{e}}rard-Marchant, Sheppard, Reddy, Weckesser, Abbasi, Gohlke, \&
  Oliphant}]{harris20}
Harris, C.~R., Millman, K.~J., van~der Walt, S.~J., {et~al.} 2020,
  \href{http://dx.doi.org/10.1038/s41586-020-2649-2}{\color{magenta}Nature},
  585, 585

\bibitem[{{Hewitt} {et~al.}(2020){Hewitt}, {Pretorius}, {Woudt}, {Tremou},
  {Miller-Jones}, {Knigge}, {Castro Segura}, {Williams}, {Fender}, {Armstrong},
  {Groot}, {Heywood}, {Horesh}, {van der Horst}, {Koerding}, {McBride},
  {Mooley}, {Rowlinson}, {Stappers}, \& {Wijers}}]{hewitt20}
{Hewitt}, D.~M., {Pretorius}, M.~L., {Woudt}, P.~A., {et~al.} 2020,
  \href{http://dx.doi.org/10.1093/mnras/staa1747}{\color{magenta}\mnras},
  \href{https://ui.adsabs.harvard.edu/abs/2020MNRAS.496.2542H}{496, 2542}

\bibitem[{{Hill} {et~al.}(2011){Hill}, {Szostek}, {Corbel}, {Camilo}, {Corbet},
  {Dubois}, {Dubus}, {Edwards}, {Ferrara}, \& {Kerr}}]{hill11}
{Hill}, A.~B., {Szostek}, A., {Corbel}, S., {et~al.} 2011,
  \href{http://dx.doi.org/10.1111/j.1365-2966.2011.18692.x}{\color{magenta}\mnras},
  \href{https://ui.adsabs.harvard.edu/abs/2011MNRAS.415..235H}{415, 235}

\bibitem[{Hunter(2007)}]{hunter07}
Hunter, J.~D. 2007,
  \href{http://dx.doi.org/10.1109/MCSE.2007.55}{\color{magenta}Comput. Sci.
  Eng.}, 9, 9

\bibitem[{Jaodand(2019)}]{jaodand19thesis}
Jaodand, A.~D. 2019, {Unravelling the nature of transitional millisecond
  pulsars}, PhD thesis, U. Amsterdam

\bibitem[{{Jaodand} {et~al.}(2021){Jaodand}, {Hern{\'a}ndez Santisteban},
  {Archibald}, {Hessels}, {Bogdanov}, {Knigge}, {Degenaar}, {Deller},
  {Scaringi}, \& {Patruno}}]{jaodand21}
{Jaodand}, A.~D., {Hern{\'a}ndez Santisteban}, J.~V., {Archibald}, A.~M.,
  {et~al.} 2021,
  \href{https://ui.adsabs.harvard.edu/abs/2021arXiv210213145J}{arXiv e-prints,
  arXiv:2102.13145}

\bibitem[{{Jonas} \& {MeerKAT Team}(2016)}]{jonas16}
{Jonas}, J. \& {MeerKAT Team}. 2016, in Proceedings of MeerKAT Science: On the
  Pathway to the SKA. 25-27 May,
  \href{https://ui.adsabs.harvard.edu/abs/2016mks..confE...1J}{1}

\bibitem[{{Joye} \& {Mandel}(2003)}]{joye03}
{Joye}, W.~A. \& {Mandel}, E. 2003, in Astronomical Society of the Pacific
  Conference Series, Vol. 295, Astronomical Data Analysis Software and Systems
  XII, ed. H.~E. {Payne}, R.~I. {Jedrzejewski}, \& R.~N. {Hook},
  \href{https://ui.adsabs.harvard.edu/abs/2003ASPC..295..489J}{489}

\bibitem[{Kendall(1938)}]{kendall38}
Kendall, M.~G. 1938,
  \href{http://dx.doi.org/10.1093/biomet/30.1-2.81}{\color{magenta}Biometrika},
  30, 30

\bibitem[{{Kenyon} {et~al.}(2018){Kenyon}, {Smirnov}, {Grobler}, \&
  {Perkins}}]{kenyon18}
{Kenyon}, J.~S., {Smirnov}, O.~M., {Grobler}, T.~L., \& {Perkins}, S.~J. 2018,
  \href{http://dx.doi.org/10.1093/mnras/sty1221}{\color{magenta}\mnras},
  \href{https://ui.adsabs.harvard.edu/abs/2018MNRAS.478.2399K}{478, 2399}

\bibitem[{{Li} {et~al.}(2020){Li}, {Strader}, {Miller-Jones}, {Heinke}, \&
  {Chomiuk}}]{li20}
{Li}, K.-L., {Strader}, J., {Miller-Jones}, J. C.~A., {Heinke}, C.~O., \&
  {Chomiuk}, L. 2020,
  \href{http://dx.doi.org/10.3847/1538-4357/ab8f28}{\color{magenta}\apj},
  \href{https://ui.adsabs.harvard.edu/abs/2020ApJ...895...89L}{895, 89}

\bibitem[{{Lindegren} {et~al.}(2021){Lindegren}, {Klioner}, {Hern{\'a}ndez},
  {Bombrun}, {Ramos-Lerate}, {Steidelm{\"u}ller}, {Bastian}, {Biermann}, {de
  Torres}, {Gerlach}, {Geyer}, {Hilger}, {Hobbs}, {Lammers}, {McMillan},
  {Stephenson}, {Casta{\~n}eda}, {Davidson}, {Fabricius}, {Gracia-Abril},
  {Portell}, {Rowell}, {Teyssier}, {Torra}, {Bartolom{\'e}}, {Clotet},
  {Garralda}, {Gonz{\'a}lez-Vidal}, {Torra}, {Abbas}, {Altmann}, {Anglada
  Varela}, {Balaguer-N{\'u}{\~n}ez}, {Balog}, {Barache}, {Becciani}, {Bernet},
  {Bertone}, {Bianchi}, {Bouquillon}, {Brown}, {Bucciarelli}, {Busonero},
  {Butkevich}, {Buzzi}, {Cancelliere}, {Carlucci}, {Charlot}, {Cioni},
  {Crosta}, {Crowley}, {del Peloso}, {del Pozo}, {Drimmel}, {Esquej}, {Fienga},
  {Fraile}, {Gai}, {Garcia-Reinaldos}, {Guerra}, {Hambly}, {Hauser},
  {Jan{\ss}en}, {Jordan}, {Kostrzewa-Rutkowska}, {Lattanzi}, {Liao}, {Licata},
  {Lister}, {L{\"o}ffler}, {Marchant}, {Masip}, {Mignard}, {Mints}, {Molina},
  {Mora}, {Morbidelli}, {Murphy}, {Pagani}, {Panuzzo}, {Pe{\~n}alosa Esteller},
  {Poggio}, {Re Fiorentin}, {Riva}, {Sagrist{\`a} Sell{\'e}s}, {Sanchez
  Gimenez}, {Sarasso}, {Sciacca}, {Siddiqui}, {Smart}, {Souami}, {Spagna},
  {Steele}, {Taris}, {Utrilla}, {van Reeven}, \& {Vecchiato}}]{lindegren21}
{Lindegren}, L., {Klioner}, S.~A., {Hern{\'a}ndez}, J., {et~al.} 2021,
  \href{http://dx.doi.org/10.1051/0004-6361/202039709}{\color{magenta}\aap},
  \href{https://ui.adsabs.harvard.edu/abs/2021A&A...649A...2L}{649, A2}

\bibitem[{Longair(2011)}]{longair11}
Longair, M.~S. 2011, High energy astrophysics, 3rd edn. (Cambridge: Cambridge
  University Press)

\bibitem[{{Mason} {et~al.}(2001){Mason}, {Breeveld}, {Much}, {Carter},
  {Cordova}, {Cropper}, {Fordham}, {Huckle}, {Ho}, {Kawakami}, {Kennea},
  {Kennedy}, {Mittaz}, {Pandel}, {Priedhorsky}, {Sasseen}, {Shirey}, {Smith},
  \& {Vreux}}]{mason01}
{Mason}, K.~O., {Breeveld}, A., {Much}, R., {et~al.} 2001,
  \href{http://dx.doi.org/10.1051/0004-6361:20000044}{\color{magenta}\aap},
  \href{https://ui.adsabs.harvard.edu/abs/2001A%26A...365L..36M}{365, L36}

\bibitem[{{Mauch} {et~al.}(2020){Mauch}, {Cotton}, {Condon}, {Matthews},
  {Abbott}, {Adam}, {Aldera}, {Asad}, {Bauermeister}, {Bennett}, {Bester},
  {Botha}, {Brederode}, {Brits}, {Buchner}, {Burger}, {Camilo}, {Chalmers},
  {Cheetham}, {de Villiers}, {de Villiers}, {Dikgale-Mahlakoana}, {du Toit},
  {Esterhuyse}, {Fadana}, {Fanaroff}, {Fataar}, {February}, {Frank},
  {Gamatham}, {Geyer}, {Goedhart}, {Gounden}, {Gumede}, {Heywood}, {Hlakola},
  {Horrell}, {Hugo}, {Isaacson}, {J{\'o}zsa}, {Jonas}, {Julie}, {Kapp},
  {Kasper}, {Kenyon}, {Kotz{\'e}}, {Kriek}, {Kriel}, {Kusel}, {Lehmensiek},
  {Loots}, {Lord}, {Lunsky}, {Madisa}, {Magnus}, {Main}, {Malan}, {Manley},
  {Marais}, {Martens}, {Merry}, {Millenaar}, {Mnyandu}, {Moeng}, {Mokone},
  {Monama}, {Mphego}, {New}, {Ngcebetsha}, {Ngoasheng}, {Ockards}, {Oozeer},
  {Otto}, {Patel}, {Peens-Hough}, {Perkins}, {Ramaila}, {Ramudzuli}, {Renil},
  {Richter}, {Robyntjies}, {Salie}, {Schollar}, {Schwardt}, {Serylak},
  {Siebrits}, {Sirothia}, {Smirnov}, {Sofeya}, {Stone}, {Taljaard}, {Tasse},
  {Theron}, {Tiplady}, {Toruvanda}, {Twum}, {van Balla}, {van der Byl}, {van
  der Merwe}, {Van Tonder}, {Wallace}, {Welz}, {Williams}, \& {Xaia}}]{mauch20}
{Mauch}, T., {Cotton}, W.~D., {Condon}, J.~J., {et~al.} 2020,
  \href{http://dx.doi.org/10.3847/1538-4357/ab5d2d}{\color{magenta}\apj},
  \href{https://ui.adsabs.harvard.edu/abs/2020ApJ...888...61M}{888, 61}

\bibitem[{{McMullin} {et~al.}(2007){McMullin}, {Waters}, {Schiebel}, {Young},
  \& {Golap}}]{mcmullin07}
{McMullin}, J.~P., {Waters}, B., {Schiebel}, D., {Young}, W., \& {Golap}, K.
  2007, in Astronomical Society of the Pacific Conference Series, Vol. 376,
  Astronomical Data Analysis Software and Systems XVI, ed. R.~A. {Shaw},
  F.~{Hill}, \& D.~J. {Bell},
  \href{https://ui.adsabs.harvard.edu/abs/2007ASPC..376..127M}{127}

\bibitem[{{Mohan} \& {Rafferty}(2015)}]{mohan15}
{Mohan}, N. \& {Rafferty}, D. 2015, {PyBDSF: Python Blob Detection and Source
  Finder}

\bibitem[{{Motta} {et~al.}(2021){Motta}, {Tremou}, {Fender}, {Eijnden},
  {Williams}, {Woudt}, \& {Miller-Jones}}]{motta21}
{Motta}, S.~E., {Tremou}, E., {Fender}, R., {et~al.} 2021, The Astronomer's
  Telegram, \href{https://ui.adsabs.harvard.edu/abs/2021ATel14659....1M}{14659,
  1}

\bibitem[{{Paduano} {et~al.}(2021){Paduano}, {Bahramian}, {Miller-Jones},
  {Kawka}, {Strader}, {Chomiuk}, {Heinke}, {Maccarone}, {Britt}, {Plotkin},
  {Shaw}, {Shishkovsky}, {Tremou}, {Tudor}, \& {Sivakoff}}]{paduano21}
{Paduano}, A., {Bahramian}, A., {Miller-Jones}, J. C.~A., {et~al.} 2021,
  \href{http://dx.doi.org/10.1093/mnras/stab1928}{\color{magenta}\mnras},
  \href{https://ui.adsabs.harvard.edu/abs/2021MNRAS.506.4107P}{506, 4107}

\bibitem[{{Papitto} {et~al.}(2019){Papitto}, {Ambrosino}, {Stella}, {Torres},
  {Coti Zelati}, {Ghedina}, {Meddi}, {Sanna}, {Casella}, {Dallilar},
  {Eikenberry}, {Israel}, {Onori}, {Piranomonte}, {Bozzo}, {Burderi},
  {Campana}, {de Martino}, {Di Salvo}, {Ferrigno}, {Rea}, {Riggio}, {Serrano},
  {Veledina}, \& {Zampieri}}]{papitto19}
{Papitto}, A., {Ambrosino}, F., {Stella}, L., {et~al.} 2019,
  \href{http://dx.doi.org/10.3847/1538-4357/ab2fdf}{\color{magenta}\apj},
  \href{https://ui.adsabs.harvard.edu/abs/2019ApJ...882..104P}{882, 104}

\bibitem[{{Papitto} \& {de Martino}(2020)}]{papitto20}
{Papitto}, A. \& {de Martino}, D. 2020,
  \href{https://ui.adsabs.harvard.edu/abs/2020arXiv201009060P}{arXiv e-prints,
  arXiv:2010.09060}

\bibitem[{{Papitto} {et~al.}(2013){Papitto}, {Ferrigno}, {Bozzo}, {Rea},
  {Pavan}, {Burderi}, {Burgay}, {Campana}, {di Salvo}, \&
  {Falanga}}]{papitto13}
{Papitto}, A., {Ferrigno}, C., {Bozzo}, E., {et~al.} 2013,
  \href{http://dx.doi.org/10.1038/nature12470}{\color{magenta}\nat},
  \href{https://ui.adsabs.harvard.edu/abs/2013Natur.501..517P}{501, 517}

\bibitem[{{Papitto} \& {Torres}(2015)}]{papitto15}
{Papitto}, A. \& {Torres}, D.~F. 2015,
  \href{http://dx.doi.org/10.1088/0004-637X/807/1/33}{\color{magenta}\apj},
  \href{https://ui.adsabs.harvard.edu/abs/2015ApJ...807...33P}{807, 33}

\bibitem[{{Parikh} {et~al.}(2019){Parikh}, {Russell}, {Wijnands},
  {Miller-Jones}, {Sivakoff}, \& {Tetarenko}}]{parikh19}
{Parikh}, A.~S., {Russell}, T.~D., {Wijnands}, R., {et~al.} 2019,
  \href{http://dx.doi.org/10.3847/2041-8213/ab2636}{\color{magenta}\apjl},
  \href{https://ui.adsabs.harvard.edu/abs/2019ApJ...878L..28P}{878, L28}

\bibitem[{{Reynolds}(1994)}]{reynolds94}
{Reynolds}, J.~E. 1994, {ATNF Memo},
  \href{http://www.atnf.csiro.au/observers/memos/d96783~1.pdf}{{AT/39.3/040}}

\bibitem[{{Robitaille} \& {Bressert}(2012)}]{robitaille12}
{Robitaille}, T. \& {Bressert}, E. 2012, {APLpy: Astronomical Plotting Library
  in Python}

\bibitem[{{Smirnov}(2011)}]{smirnov11}
{Smirnov}, O.~M. 2011,
  \href{http://dx.doi.org/10.1051/0004-6361/201016082}{\color{magenta}\aap},
  \href{https://ui.adsabs.harvard.edu/abs/2011A&A...527A.106S}{527, A106}

\bibitem[{Spearman(1904)}]{spearman04}
Spearman, C. 1904, The American Journal of Psychology, 15, 15

\bibitem[{{Stappers} {et~al.}(2014){Stappers}, {Archibald}, {Hessels}, {Bassa},
  {Bogdanov}, {Janssen}, {Kaspi}, {Lyne}, {Patruno}, \&
  {Tendulkar}}]{stappers14}
{Stappers}, B.~W., {Archibald}, A.~M., {Hessels}, J.~W.~T., {et~al.} 2014,
  \href{http://dx.doi.org/10.1088/0004-637X/790/1/39}{\color{magenta}\apj},
  \href{https://ui.adsabs.harvard.edu/abs/2014ApJ...790...39S}{790, 39}

\bibitem[{Stella \& Angelini(1992)}]{stella92}
Stella, L. \& Angelini, L. 1992, XRONOS: A Timing Analysis Software Package,
  ed. V.~Di~Ges{\`u}, L.~Scarsi, R.~Buccheri, P.~Crane, M.~C. Maccarone, \&
  H.~U. Zimmermann (Boston, MA: Springer US), 59--64

\bibitem[{{Strader} {et~al.}(2021){Strader}, {Swihart}, {Urquhart}, {Chomiuk},
  {Aydi}, {Bahramian}, {Kawash}, {Sokolovsky}, {Tremou}, \&
  {Udalski}}]{strader21}
{Strader}, J., {Swihart}, S.~J., {Urquhart}, R., {et~al.} 2021,
  \href{http://dx.doi.org/10.3847/1538-4357/ac0b47}{\color{magenta}\apj},
  \href{https://ui.adsabs.harvard.edu/abs/2021ApJ...917...69S}{917, 69}

\bibitem[{{Str{\"u}der} {et~al.}(2001){Str{\"u}der}, {Briel}, {Dennerl},
  {Hartmann}, {Kendziorra}, {Meidinger}, {Pfeffermann}, {Reppin}, {Aschenbach},
  \& {Bornemann}}]{struder01}
{Str{\"u}der}, L., {Briel}, U., {Dennerl}, K., {et~al.} 2001,
  \href{http://dx.doi.org/10.1051/0004-6361:20000066}{\color{magenta}\aap},
  \href{https://ui.adsabs.harvard.edu/abs/2001A&A...365L..18S}{365, L18}

\bibitem[{{Tanaka} \& {Takahara}(2010)}]{Tanaka2010}
{Tanaka}, S.~J. \& {Takahara}, F. 2010,
  \href{http://dx.doi.org/10.1088/0004-637X/715/2/1248}{\color{magenta}\apj},
  \href{https://ui.adsabs.harvard.edu/abs/2010ApJ...715.1248T}{715, 1248}

\bibitem[{{Tasse} {et~al.}(2018){Tasse}, {Hugo}, {Mirmont}, {Smirnov},
  {Atemkeng}, {Bester}, {Hardcastle}, {Lakhoo}, {Perkins}, \&
  {Shimwell}}]{tasse18}
{Tasse}, C., {Hugo}, B., {Mirmont}, M., {et~al.} 2018,
  \href{http://dx.doi.org/10.1051/0004-6361/201731474}{\color{magenta}\aap},
  \href{https://ui.adsabs.harvard.edu/abs/2018A&A...611A..87T}{611, A87}

\bibitem[{{Torres} {et~al.}(2014){Torres}, {Cillis}, {Mart{\'\i}n}, \& {de
  O{\~n}a Wilhelmi}}]{Torres14}
{Torres}, D.~F., {Cillis}, A., {Mart{\'\i}n}, J., \& {de O{\~n}a Wilhelmi}, E.
  2014, \href{http://dx.doi.org/10.1016/j.jheap.2014.02.001}{\color{magenta}J.
  High Energy Astrophys.},
  \href{https://ui.adsabs.harvard.edu/abs/2014JHEAp...1...31T}{1, 31}

\bibitem[{{Tremou} {et~al.}(2020){Tremou}, {Corbel}, {Fender}, {Woudt},
  {Miller-Jones}, {Motta}, {Heywood}, {Armstrong}, {Groot}, {Horesh}, {van der
  Horst}, {Koerding}, {Mooley}, {Rowlinson}, \& {Wijers}}]{tremou20}
{Tremou}, E., {Corbel}, S., {Fender}, R.~P., {et~al.} 2020,
  \href{http://dx.doi.org/10.1093/mnrasl/slaa019}{\color{magenta}\mnras},
  \href{https://ui.adsabs.harvard.edu/abs/2020MNRAS.493L.132T}{493, L132}

\bibitem[{{Turner} {et~al.}(2001){Turner}, {Abbey}, {Arnaud}, {Balasini},
  {Barbera}, {Belsole}, {Bennie}, {Bernard}, {Bignami}, {Boer}, {Briel},
  {Butler}, {Cara}, {Chabaud}, {Cole}, {Collura}, {Conte}, {Cros}, {Denby},
  {Dhez}, {Di Coco}, {Dowson}, {Ferrando}, {Ghizzardi}, {Gianotti}, {Goodall},
  {Gretton}, {Griffiths}, {Hainaut}, {Hochedez}, {Holland}, {Jourdain},
  {Kendziorra}, {Lagostina}, {Laine}, {La Palombara}, {Lortholary}, {Lumb},
  {Marty}, {Molendi}, {Pigot}, {Poindron}, {Pounds}, {Reeves}, {Reppin},
  {Rothenflug}, {Salvetat}, {Sauvageot}, {Schmitt}, {Sembay}, {Short},
  {Spragg}, {Stephen}, {Str{\"u}der}, {Tiengo}, {Trifoglio}, {Tr{\"u}mper},
  {Vercellone}, {Vigroux}, {Villa}, {Ward}, {Whitehead}, \& {Zonca}}]{turner01}
{Turner}, M.~J.~L., {Abbey}, A., {Arnaud}, M., {et~al.} 2001,
  \href{http://dx.doi.org/10.1051/0004-6361:20000087}{\color{magenta}\aap},
  \href{https://ui.adsabs.harvard.edu/abs/2001A%26A...365L..27T}{365, L27}

\bibitem[{{van den Eijnden} {et~al.}(2020){van den Eijnden}, {Degenaar},
  {Russell}, {Buisson}, {Altamirano}, {Armas Padilla}, {Bahramian}, {Castro
  Segura}, {Fogantini}, {Heinke}, {Maccarone}, {Maitra}, {Miller-Jones},
  {Mu{\~n}oz-Darias}, {{\"O}zbey Arabac{\i}}, {Russell}, {Shaw}, {Sivakoff},
  {Tetarenko}, {Vincentelli}, \& {Wijnand s}}]{vandene20}
{van den Eijnden}, J., {Degenaar}, N., {Russell}, T.~D., {et~al.} 2020,
  \href{http://dx.doi.org/10.1093/mnras/staa1704}{\color{magenta}\mnras},
  \href{https://ui.adsabs.harvard.edu/abs/2020MNRAS.496.4127V}{496, 4127}

\bibitem[{{van den Eijnden} {et~al.}(2021){van den Eijnden}, {Degenaar},
  {Russell}, {Wijnands}, {Bahramian}, {Miller-Jones}, {Hern{\'a}ndez
  Santisteban}, {Gallo}, {Atri}, {Plotkin}, {Maccarone}, {Sivakoff}, {Miller},
  {Reynolds}, {Russell}, {Maitra}, {Heinke}, {Armas Padilla}, \&
  {Shaw}}]{vandene21}
{van den Eijnden}, J., {Degenaar}, N., {Russell}, T.~D., {et~al.} 2021,
  \href{http://dx.doi.org/10.1093/mnras/stab1995}{\color{magenta}\mnras},
  \href{https://ui.adsabs.harvard.edu/abs/2021MNRAS.507.3899V}{507, 3899}

\bibitem[{{Veledina} {et~al.}(2019){Veledina}, {N{\"a}ttil{\"a}}, \&
  {Beloborodov}}]{veledina19}
{Veledina}, A., {N{\"a}ttil{\"a}}, J., \& {Beloborodov}, A.~M. 2019,
  \href{http://dx.doi.org/10.3847/1538-4357/ab44c6}{\color{magenta}\apj},
  \href{https://ui.adsabs.harvard.edu/abs/2019ApJ...884..144V}{884, 144}

\bibitem[{Virtanen {et~al.}(2020)Virtanen, Gommers, Oliphant, Haberland, Reddy,
  Cournapeau, Burovski, Peterson, Weckesser, Bright, {van der Walt}, Brett,
  Wilson, Millman, Mayorov, Nelson, Jones, Kern, Larson, Carey, Polat, Feng,
  Moore, {VanderPlas}, Laxalde, Perktold, Cimrman, Henriksen, Quintero, Harris,
  Archibald, Ribeiro, Pedregosa, {van Mulbregt}, \& {SciPy 1.0
  Contributors}}]{scipy20}
Virtanen, P., Gommers, R., Oliphant, T.~E., {et~al.} 2020,
  \href{http://dx.doi.org/10.1038/s41592-019-0686-2}{\color{magenta}Nat.
  Methods}, \href{https://rdcu.be/b08Wh}{17, 261}

\bibitem[{{Williams} {et~al.}(2020){Williams}, {Motta}, {Fender}, {Bright},
  {Heywood}, {Tremou}, {Woudt}, {Buckley}, {Corbel}, {Coriat}, {Joseph},
  {Rhodes}, {Sivakoff}, \& {van der Horst}}]{williams20}
{Williams}, D.~R.~A., {Motta}, S.~E., {Fender}, R., {et~al.} 2020,
  \href{http://dx.doi.org/10.1093/mnrasl/slz152}{\color{magenta}\mnras},
  \href{https://ui.adsabs.harvard.edu/abs/2020MNRAS.491L..29W}{491, L29}

\bibitem[{{Wilms} {et~al.}(2000){Wilms}, {Allen}, \& {McCray}}]{wilms00}
{Wilms}, J., {Allen}, A., \& {McCray}, R. 2000,
  \href{http://dx.doi.org/10.1086/317016}{\color{magenta}\apj},
  \href{http://adsabs.harvard.edu/abs/2000ApJ...542..914W}{542, 914}

\bibitem[{{Xie} {et~al.}(2020){Xie}, {Yan}, \& {Wu}}]{xie20}
{Xie}, F.-G., {Yan}, Z., \& {Wu}, Z. 2020,
  \href{http://dx.doi.org/10.3847/1538-4357/ab711f}{\color{magenta}\apj},
  \href{https://ui.adsabs.harvard.edu/abs/2020ApJ...891...31X}{891, 31}

\end{thebibliography}

\begin{appendix}
\section{Direction-dependent calibration and peeling of the \meer\ data}
\label{sec:rime}
Direction-dependent calibration and peeling of the \meer\ data was accomplished by solving the following radio interferometer measurement equation (RIME):
\begin{multline}
\label{eq:rime}
V_{pq}(t)= G_p(t)\Bigg\{\mathzapf{F}\big[M_{{\rm DIE}}\big]+\\\sum_{l,m}^{{\rm dir}} \Delta_p(\nu,t,l,m)\mathzapf{F}\big[M(l,m)\big]\Delta_q^H(\nu,t,l,m)\Bigg\}G_q^H(t)~.   
\end{multline}
The term on the left-hand side $V_{pq}$ represents the visibility observed by a baseline formed by a pair of antennas $p$ and $q$, where $p,q \in [1,...,N_A]$ and $N_A$ is the total number of antennas in the interferometer. Eq.\,(\ref{eq:rime}) simultaneously solves for the temporal electronic gains $G_p$ and $G_q$ towards all the flux in the field unaffected by direction-dependent errors ($\mathzapf{F}[M_{{\rm DIE}}$]), as well as for the direction-dependent differential gains $\Delta_p$ and $\Delta_q$ for a number of sky regions in the field (e.g. polygons) tagged by the user ($l$, $m$ are the `direction cosines' toward the sky; the superscript `$H$' denotes the Hermitian transpose; see e.g. \citealt{smirnov11} for more details on the RIME formalism). The flux contained within each of these regions ($\mathzapf{F}[M(l,m)$]) is predicted through targeted faceting from the sky model produced by \textsc{DDFacet} \citep{tasse18}.
This approach of simultaneous calibration and peeling taking into account all available flux limits the suppression of unmodelled sky flux that is typically seen in traditional phase-steering and peeling methods using just a model of the source affected by direction-dependent calibration errors.
\end{appendix}

\listofobjects 
 
\end{document}